\documentclass[english,journal=jctcce,manuscript=article,email=true,etalmode=truncate,maxauthors=0]{achemso}

  \usepackage{geometry} 
  \usepackage{stackengine}
  \usepackage{titlesec}

  \usepackage[T1]{fontenc} 
  \usepackage{amsmath}     
  \usepackage{amssymb}
  \usepackage{bm}
  \usepackage{mathrsfs}
  \usepackage{etoolbox}
  \usepackage{environ}
  \usepackage[version=4]{mhchem}
  \usepackage{booktabs}
  \usepackage{adjustbox}
  \usepackage{hhline}

  \usepackage{setspace}    
  \usepackage{float}       
  \newfloat{chart}{htbp}{loc}
  \newfloat{graph}{htbp}{loh}
  \newfloat{scheme}{htbp}{los}

  \usepackage{graphicx}    

  \usepackage[english]{babel}
  \usepackage{array}

  \usepackage{subcaption}  
  \usepackage{outlines}    
  \usepackage[normalem]{ulem} 

  \usepackage[numbers,sort&compress,super]{natbib}
  \usepackage{natmove}

  \usepackage{cancel}      
  \usepackage{pdfpages}  
  \usepackage[font=scriptsize,labelfont=bf]{caption}

  \usepackage{hyperref} 
  \usepackage{cleveref}	
  	\crefname{figure}{Figure}{Figures}
  	\crefname{table}{Table}{Tables}
  	\crefname{equation}{Eq.}{Eqs.}
  	\crefname{section}{Section}{Sections}
  	\crefname{subsection}{Section}{Sections}
  	\crefname{subsubsection}{Section}{Sections}
  	\crefname{algorithm}{Algorithm}{Algorithms}

  \usepackage{tikz}

    \usepackage[ruled,vlined]{algorithm2e}
  
  \newcommand*{\bigO}[1]{\ensuremath{\mathcal{O}({#1})}}

  \usepackage{todonotes}

  \author{Karl Pierce}
  \author{Varun Rishi}
  \author{Edward F. Valeev}
  \email{efv@vt.edu}
  \affiliation{ Department of Chemistry, Virginia Tech, Blacksburg, Virginia 24061, U.S.A. }
  \title{Robust approximation of tensor networks: application to grid-free tensor factorization of the Coulomb interaction}

\begin{document}

  \tikzstyle{term}=[circle,font=\bfseries\sffamily, inner sep=0pt, outer sep=0pt]
  \tikzstyle{circ}=[circle,draw=#1,text=#1,font=\bfseries\sffamily]
  \tikzstyle{circ0}=[circle, inner sep=2pt , draw=#1, text=#1, font=\bfseries\sffamily]
  \tikzstyle{meet}=[circle,fill,font=\bfseries\sffamily,inner sep=0pt, outer sep=0pt]
  \tikzstyle{line}=[#1]
  \tikzstyle{rline}=[bend right,#1]
  \tikzstyle{lline}=[bend left,#1]

\begin{abstract}
  Approximation of a tensor network by approximating (e.g., factorizing) one or more of its constituent tensors can be improved by canceling the leading-order error due to the constituents' approximation. The utility of such robust approximation is demonstrated for robust canonical polyadic (CP) approximation of a (density-fitting) factorized 2-particle Coulomb interaction tensor. The resulting algebraic (grid-free) approximation for the Coulomb tensor, closely related to the factorization appearing in pseudospectral and tensor hypercontraction approaches, is efficient and accurate, with significantly reduced rank compared to the naive (non-robust) approximation. Application of the robust approximation to the particle-particle ladder term in the coupled-cluster singles and doubles reduces the size complexity from \bigO{N^6} to \bigO{N^5} with robustness ensuring negligible errors in chemically-relevant energy differences using CP ranks approximately equal to the size of the density-fitting basis.
\end{abstract}

\section{Introduction}

  Numerical approximation of the (matrix elements of the) Hamiltonian is a ubiquitous strategy for decreasing the cost and complexity of
  quantum simulation of, e.g., electronic structure in both real space and spectral representations.
  Examples in spectral representations include density fitting 
  (DF: also referred to in quantum chemistry as the
  resolution-of-the-identity (RI), in global\cite{Whitten1973,Vahtras1993} and local\cite{Scuseria1999,Saebo1993,Hampel1996}), the pseudospectral\cite{Friesner:1985bo,Friesner1986,Langlois1990,Ringnalda1990,Friesner1991,Martinez1995,Martinez1992,Martinez1994,Ko2008,Martinez1993} (PS) approach, Cholesky decomposition (CD),\cite{Beebe1977,Lowdin1965,Lowdin2009,Folkestad2019} the fast multipole method (FMM),\cite{White1994,Burant1996,Rudberg2006} tensor hypercontraction (THC),\cite{Hohenstein2012,Parrish2012,Hohenstein2012a,Hohenstein2013,Parrish2014,Shenvi2013,Schutski2017,Parrish2019,Lee2019} the canonical polyadic  (CP) decomposition (also known as CANDECOMP/PARAFAC\cite{Carroll1970,Harshman1970}),
  \cite{Benedikt2011,Benedikt2013a,Benedikt2013,Hummel2017,Bohm2016,Chinnamsetty2007,Khoromskij2009} and many others.\cite{Lewis2016,Bischoff2011,Fusti-Molnar2002,Dutta2016,Izsak2013,Izsak2012,Petrenko2011,Izsak2011,Kossmann2010,Neese2009c}
  These approaches can be coarsely classified as (a) abstract (algebraic) approximations of the Hamiltonian tensor (e.g., CD, CP, {\em global} DF, {\em algebraic} FMM\cite{Sun:2001fb,Borm:2003hk}),
  and (b) approximations that utilize physical context (e.g., use of grids in pseudospectral and THC, domain decomposition in FMM and {\em local} DF).

  It is common to wish to approximate tensors in a tensor {\em network}. In such a case, it may be possible to construct a better network approximation to the original tensor network than obtained by approximating the individual tensors in the network.
  Inspired by these basic observations we consider the {\em robust}\footnote{In this work, the term ``robust''
  mirrors its use in the discussion of fitting in quantum chemistry\cite{Dunlap:2000iw} rather than referring to
  the robust approximation of individual tensors.\cite{Goldfarb:2014dy}} approximation of tensor networks, in which the leading-order error due to the approximation of the network constituents is cancelled.
  Here, we demonstrate the utility of the idea by constructing a robust CP (rCP) approximation for a simple network of two order-3 tensors
  obtained by the DF-factorization of the 2-particle Coulomb interaction tensor.
  Unlike DF-factorization alone, the rCP-DF decomposition reduces the complexity of the ladder-type diagrams in many-body electronic structure methods.
  The robustness of the approximation ensures a favorable prefactor; in this work, cost savings are observed for systems with as few as $3$ atoms, as demonstrated for
  the particle-particle ladder (PPL) diagram in the coupled cluster method with single and double excitations (CCSD).

  The rest of manuscript is organized as follows. In \cref{sec:theory} of this paper we introduce the idea of robust approximation of tensor networks, use it to construct an efficient algebraic approximation to a 2-particle interaction tensor, and discuss how to utilize the proposed factorization to evaluate the particle-particle ladder (PPL) diagram with reduced complexity. \Cref{sec:comp_det} describes the details of the computational experiments. \cref{sec:results} compares the performances of non-robust and robust approximations applied to the CCSD PPL  diagram using standard benchmark sets of noncovalent interaction energies and reaction energies. \cref{sec:discuss} summarizes our findings and discusses other possible applications of the idea.

\section{Formalism} \label{sec:theory}
  \subsection{Robust Approximation of Tensor Networks}
  \label{sec:robusttnapp}
     Consider a tensor network composed of a sequence of tensors, $\{\mathcal{T}_1 \dots \mathcal{T}_k\} \equiv \{\mathcal{T}_i\}, i = 1\dots k$. For our purposes the network can have arbitrary topology,
     it does not even need to be connected. Our objective is to minimize the error in the network due to replacing tensors $\mathcal{T}_i$ by their approximants 
     $\hat{\mathcal{T}}_i$. Assuming that the approximation error in each tensor,
     \begin{align}
     \label{eq:delta}
      \delta_i \equiv \mathcal{T}_i - \hat{\mathcal{T}}_i,
     \end{align}
     is ``small'', i.e., $||\delta_i|| = \bigO{\epsilon}$,
     the tensor network can be accurately represented in terms of tensor approximants by including terms linear in the error:
     \begin{align}
       \label{eq:approx}
       \{\mathcal{T}_1 \dots \mathcal{T}_k\} = \{\hat{\mathcal{T}}_1 \dots \hat{\mathcal{T}}_k\} +  \sum_j \{\hat{\mathcal{T}}_1 \dots \hat{\mathcal{T}}_{j-1} \delta_j \hat{\mathcal{T}}_{j+1} \dots \hat{\mathcal{T}}_k \} + \bigO{\epsilon^2}.
     \end{align}
     Note that the {\em naive} approximation of the network, given by the first term on the right-hand side, is only accurate to $\bigO{\epsilon}$.
     A {\em robust} approximation, accurate to $\bigO{\epsilon^2}$, is obtained by plugging \cref{eq:delta} into \cref{eq:approx}:
     \begin{align}
       \label{eq:approx2}
       \{\mathcal{T}_1 \dots \mathcal{T}_k\} = (1 - k) \{\hat{\mathcal{T}}_1 \dots \hat{\mathcal{T}}_k\} +  \sum_j \{\hat{\mathcal{T}}_1 \dots \hat{\mathcal{T}}_{j-1} \mathcal{T}_j \hat{\mathcal{T}}_{j+1} \dots \hat{\mathcal{T}}_k \} + \bigO{\epsilon^2}.
     \end{align}
     Clearly, the robust approximation is only applicable to tensor networks, not individual tensors.

     In the context of numerical tensor approximations, the robust approximation has enjoyed a long use by the electronic structure community.\cite{Dunlap:2000iw,Reine2008,Izsak2013,Merlot2013}
     Despite its simplicity and/or apparent lack of novelty, in the context of tensor computation the the idea has potentially significant unexplored
     utility. Its utility came as a real surprise to us when we stumbled on its novel application, described below.
     
  \subsection{Robust approximation of factorized 2-particle interaction tensor}

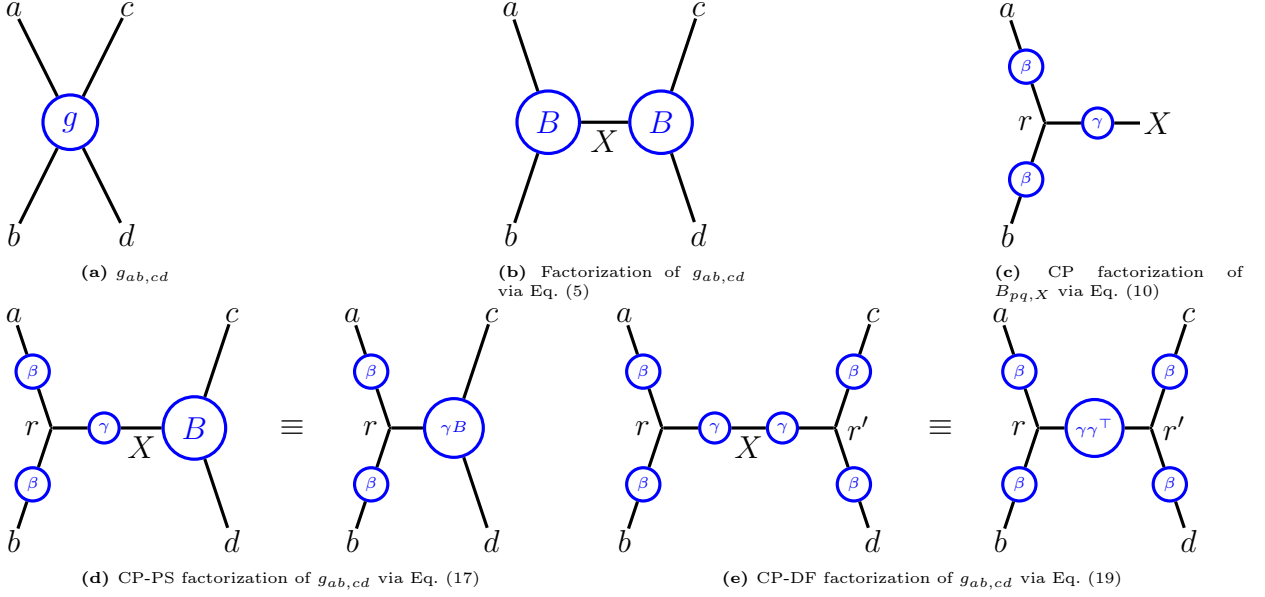
\begin{figure}[ht]
      \begin{subfigure}[t]{0.20\textwidth}
        \begin{tikzpicture}[line width=1.2pt]

          \path
          (0,0)       node[term] (B1) {$a$} 
          (1.5,0)       node[term] (B2) {$c$}
          (0.75,-1.5)    node[circ=blue] (g) {$g$}
          (0,-3)      node[term] (K1) {$b$} 
          (1.5,-3)      node[term] (K2) {$d$}
          ;

          \path
          (g) edge [line] (K1)
          (g) edge [line] (K2)
          (g) edge [line] (B1)
          (g) edge [line] (B2)
          ;
        \end{tikzpicture}
        \caption{$g_{ab,cd}$}
        \label{fig:g}
      \end{subfigure} \hfill
      \begin{subfigure}[t]{0.20\textwidth}
        \begin{tikzpicture}[line width=1.2pt]

          \path
          (0,0)       node[term] (B1) {$a$} 
          (2.5,0)       node[term] (B2) {$c$}
          (0.5,-1.5)    node[circ=blue] (b1) {$B$}
          (2.0,-1.5)    node[circ=blue] (b2) {$B$}
          (1.25,-1.75)    node[term] (2) {$X$}
          (0,-3)      node[term] (K1) {$b$} 
          (2.5,-3)      node[term] (K2) {$d$}
          ;

          \path
          (b1) edge [line] (K1)
          (b2) edge [line] (K2)
          (b1) edge [line] (B1)
          (b2) edge [line] (B2)
          (b1) edge [line] (b2)
          ;
        \end{tikzpicture}
        \caption{Factorization of $g_{ab,cd}$ via \cref{eq:df_approx1}}
        \label{fig:df_approx1}
      \end{subfigure} \hfill
      \begin{subfigure}[t]{0.20\textwidth}
        \begin{tikzpicture}[line width=1.2pt]

          \path
          (0,0)         node[term] (B1) {$a$}
          (0.5,-1.5)    node[meet] (b1) {}
          (0,-3)        node[term] (K1) {$b$}
          (2.0,-1.5)    node[term] (2) {$X$}
          (0.25,-0.75)  node[circ0=blue] (beta1) {{\tiny $\beta$}}
          (0.25,-2.25)  node[circ0=blue] (kappa1) {{\tiny $\beta$}}
          (1.2,-1.50)   node[circ0=blue] (gamma1) {{\tiny $\gamma$}}
          ;

          \draw (b1) node[left]{$r$};

          \path
          (beta1) edge [line] (B1)
          (b1) edge [line] (beta1)
          (kappa1) edge [line] (K1)
          (b1) edge [line] (kappa1)
          (b1) edge [line] (gamma1)
          (gamma1) edge [line] (2)
          ;
        \end{tikzpicture}

        \caption{CP factorization of $B_{pq,X}$ via \cref{eq:B_cp3_symm}}
        \label{fig:B_cp3}
      \end{subfigure} \hfill
      \begin{subfigure}[t]{0.45\textwidth}
        \begin{tikzpicture}[line width=1.2pt]

          \path
          (0,0)         node[term] (B1) {$a$}
          (0.5,-1.5)    node[meet] (b1) {}
          (0,-3)        node[term] (K1) {$b$}
          (1.7,-1.75)    node[term] (2) {$X$}
          (0.25,-0.75)  node[circ0=blue] (beta1) {{\tiny $\beta$}}
          (0.25,-2.25)  node[circ0=blue] (kappa1) {{\tiny $\beta$}}
          (1.2,-1.50)   node[circ0=blue] (gamma1) {{\tiny $\gamma$}}
          ;

          \draw (b1) node[left]{$r$};

          \path
          (beta1) edge [line] (B1)
          (b1) edge [line] (beta1)
          (kappa1) edge [line] (K1)
          (b1) edge [line] (kappa1)
          (b1) edge [line] (gamma1)
          ;
          
          \path
          (2.9,0)       node[term] (B2) {$c$}
          (2.4,-1.5)    node[circ=blue] (b2) {$B$}
          (2.9,-3)      node[term] (K2) {$d$}
          ;

          \path
          (b2) edge [line] (B2)
          (b2) edge [line] (K2)
          (b2) edge [line] (gamma1)
          ;

          \path
          (3.7,-1.5)         node[term] (equiv) {$\equiv$}
          ;

          \path
          (4.5,0)       node[term] (B1) {$a$}
          (5,-1.5)    node[meet] (b1) {}
          (4.5,-3)      node[term] (K1) {$b$}
          (4.75,-0.75)    node[circ0=blue] (beta1) {{\tiny $\beta$}}
          (4.75,-2.25)    node[circ0=blue] (kappa1) {{\tiny $\beta$}}
          (5.85,-1.5)    node[circ=blue] (b2) {{\tiny $\gamma B$}}
          (6.35,0)       node[term] (B2) {$c$}
          (6.35,-3)      node[term] (K2) {$d$}
          ;

          \draw (b1) node[left]{$r$};

          \path
          (beta1) edge [line] (B1)
          (b1) edge [line] (beta1)
          (kappa1) edge [line] (K1)
          (b1) edge [line] (kappa1)
          (b1) edge [line] (b2)
          (b2) edge [line] (K2)
          (b2) edge [line] (B2)
          ;
        \end{tikzpicture}
        \caption{CP-PS factorization of $g_{ab,cd}$ via \cref{eq:g_cp_ps}}
        \label{fig:g_cp_ps}
      \end{subfigure} \hfill
      \begin{subfigure}[t]{0.50\textwidth}
        \begin{tikzpicture}[line width=1.2pt]

          \path
          (0,0)         node[term] (B1) {$a$}
          (0.5,-1.5)    node[meet] (b1) {}
          (0,-3)        node[term] (K1) {$b$}
          (1.65,-1.75)    node[term] (2) {$X$}
          (0.25,-0.75)  node[circ0=blue] (beta1) {{\tiny $\beta$}}
          (0.25,-2.25)  node[circ0=blue] (kappa1) {{\tiny $\beta$}}
          (1.2,-1.50)   node[circ0=blue] (gamma1) {{\tiny $\gamma$}}
          ;

          \draw (b1) node[left]{$r$};

          \path
          (beta1) edge [line] (B1)
          (b1) edge [line] (beta1)
          (kappa1) edge [line] (K1)
          (b1) edge [line] (kappa1)
          (b1) edge [line] (gamma1)
          ;
          
          \path
          (3.3,0)       node[term] (B2) {$c$}
          (2.8,-1.5)    node[meet] (b2) {}
          (3.3,-3)      node[term] (K2) {$d$}
          (3.05,-0.75)  node[circ0=blue] (beta2) {{\tiny $\beta$}}
          (3.05,-2.25)  node[circ0=blue] (kappa2) {{\tiny $\beta$}}
          (2.1,-1.50)   node[circ0=blue] (gamma2) {{\tiny $\gamma$}}
          ;

          \draw (b2) node[right]{$r'$};

          \path
          (beta2) edge [line] (B2)
          (b2) edge [line] (beta2)
          (kappa2) edge [line] (K2)
          (b2) edge [line] (kappa2)
          (b2) edge [line] (gamma2)
          (gamma2) edge [line] (gamma1)
          ;

          \path
          (4.2,-1.5)         node[term] (equiv) {$\equiv$}
          ;

          \path
          (5,0)       node[term] (B1) {$a$}
          (5.5,-1.5)    node[meet] (b1) {}
          (5,-3)      node[term] (K1) {$b$}
          (5.25,-0.75)    node[circ0=blue] (beta1) {{\tiny $\beta$}}
          (5.25,-2.25)    node[circ0=blue] (kappa1) {{\tiny $\beta$}}
          (6.25,-1.5)    node[circ0=blue] (gg) {{\tiny $\gamma\gamma^\top$}}
          (7.5,0)       node[term] (B2) {$c$}
          (7.0,-1.5)    node[meet] (b2) {}
          (7.5,-3)      node[term] (K2) {$d$}
          (7.25,-0.75)    node[circ0=blue] (beta2) {{\tiny $\beta$}}
          (7.25,-2.25)    node[circ0=blue] (kappa2) {{\tiny $\beta$}}
          ;

          \draw (b1) node[left]{$r$};
          \draw (b2) node[right]{$r'$};

          \path
          (beta1) edge [line] (B1)
          (b1) edge [line] (beta1)
          (kappa1) edge [line] (K1)
          (b1) edge [line] (kappa1)
          (b1) edge [line] (gg)
          (gg) edge [line] (b2)
          (beta2) edge [line] (B2)
          (b2) edge [line] (beta2)
          (kappa2) edge [line] (K2)
          (b2) edge [line] (kappa2)
          ;
        \end{tikzpicture}
        \caption{CP-DF factorization of $g_{ab,cd}$ via \cref{eq:g_cp_df}}
        \label{fig:g_cp_df}
      \end{subfigure} \hfill

      \caption{Graphical representation of the 2-particle interaction tensor (\cref{eq:gabcd}) and factorizations thereof considered in this work.}
      \label{fig:tn}
    \end{figure}

    Consider tensor representation of a 
    2-particle interaction\footnote{In this work we only consider Coulomb interactions using the Poisson kernel: $g({\bf r}_1,{\bf r}_2) \equiv |{\bf r}_1-{\bf r}_2|^{-1}$; extension to other multiplicative and non-multiplicative kernels is
    straightforward.} in a generic basis of size $n$:
     \begin{equation} \label{eq:gabcd}
        g_{ab,cd} \equiv \iint \phi_a^*({\bf r}_1) \phi_b({\bf r}_1) g({\bf r}_1,{\bf r}_2) \phi_c^*({\bf r}_2) \phi_d({\bf r}_2) \, d{\bf r}_1 \, d{\bf r}_2.
      \end{equation}
    The comma separator between indices defines the default {\em matricization}; namely, matrix {\bf O} will refer to the matricized form of tensor $O$, with
    element $O_{p_1 p_2 \dots,q_1 q_2 \dots}$ located in row $p_1 p_2 \dots$ and column $q_1 q_2 \dots$ of the matrix.
    It is also useful to convey tensor expressions diagrammatically; in Penrose notation tensor $g$ is represented as a single node (\cref{fig:g}).

    To efficiently approximate $g$, it is important to retain the analytic properties, such as symmetries and positivity.
    In this work, specifically, we must consider the properties of the Poisson kernel, $g({\bf r}_1,{\bf r}_2) = |{\bf r}_1-{\bf r}_2|^{-1}$, which is ``positive'' in both 2-particle and 1-particle senses, i.e.,
    both $\hat{g}_2 f({\bf r}_1,{\bf r}_2) \equiv g({\bf r}_1,{\bf r}_2) \times f({\bf r}_1,{\bf r}_2)$ and $\hat{g}_1 f({\bf r}_1) \equiv \int g({\bf r}_1,{\bf r}_2) f({\bf r}_2) d{\bf r}_2$, respectively,
    are positive definite operators.
    
    For positive-definite kernels, the tensor $g$ can be factorized into a symmetric form,
     \begin{equation} \label{eq:df_approx1}
        g_{ab,cd} \approx \sum_X B_{ab,X} B_{cd,X},
      \end{equation}
    which, in its matrix form, is recognized as the ubiquitous, symmetric particle-wise factorization
     \begin{equation} \label{eq:df_approx2}
        \mathbf{g} \approx \mathbf{B} \mathbf{B}^\top.
      \end{equation}
    Such ``generalized square root'' factorization is not unique. One way to compute the factorization efficiently is 
    by a (rank-revealing) Cholesky decomposition (CD);\cite{Folkestad2019}
    for any finite precision the CD rank (i.e., the number of columns of $\mathbf{B}$) is \bigO{n}.
    Another way to compute this symmetric factorization is via DF, where
      \begin{align}
        B_{ab,X} = C_{ab,Y} \left(\mathbf{G}^{1/2}\right)_{Y,X},
      \end{align}
     the fitting coefficients $C_{pq,Y}$ are determined by weighted least-squares fitting,\cite{Whitten1973,Vahtras1993,Scuseria1999,Saebo1993,Hampel1996} typically,
    using the Coulomb ``metric'':
      \begin{align}
        \left(\mathbf{G}\right)_{X,Y} \equiv \iint \phi_X(1) g(1,2) \phi_Y(2) \, d1 \, d2,
      \end{align}
    and the square root of $\mathbf{G}$ is defined by \cref{eq:df_approx2}, rather than the conventional, {\em principal} square root.
    The size of the fitting basis $\{ \phi_X \}$, denoted here by $X$, is in practice proportional to $n$.

    For large systems CD and DF approaches lead to sparse $\mathbf{B}$, however, in large basis sets the onset of sparsity can be slow and thus difficult to exploit.
    Hence, it may be worthwhile to seek more general {\em data sparsity} in $\mathbf{B}$ by further factorization.
    For example, consider the approximate CP factorization of $\mathbf{B}$:
      \begin{align}
        \label{eq:B_cp3}
        B_{ab,X} \approx \sum_r^R \beta_{a,r} \kappa_{b,r} \gamma_{X,r}
      \end{align}
    For real basis functions $g_{ab,cd}$ and, hence, $B_{ab,X}$ are
    symmetric with respect to the $a \leftrightarrow b$ permutation; this symmetry is ensured automatically if $\kappa_{b,r} \equiv \beta_{b,r}$, or
      \begin{align}
        \label{eq:B_cp3_symm}
        B_{ab,X} \approx \hat{B}_{ab,X} \equiv \sum_r^R \beta_{a,r} \beta_{b,r} \gamma_{X,r}
      \end{align}
    It is well known\cite{Hastad1990,Hillar:2013cx} that (aside from trivial examples) finding the exact CP rank $R$ is hard,
    but there are efficient ways to construct such approximations for a fixed CP rank, $R$.\cite{Acar2009,Sorber2013,Phan2013}

    Tensor factorization of Coulomb interaction \cref{eq:gabcd} that utilizes CP topology have been long employed in electronic structure.
    This is due to the natural connection between CP factorization and quadrature approximation for
    an integral over a product of three or more factors. Most relevant for our purposes is
    Friesner's pioneering use of a pseudospectral (PS) method (PS methods are also known as discrete variable representation [DVR] methods)
    to solve the Hartree-Fock equations for electrons.\cite{Friesner:1985bo}
    His work led to the pseudospectral family of methods\cite{Friesner:1985bo,Ko2008,Martinez1994,Ringnalda1990,Langlois1990,Friesner1986,Martinez1993,Martinez1995,Martinez1992} which approximate
    Coulomb integrals using a numerical quadrature over one electron.
    This quadrature approximation is also employed in
    the COSX method\cite{Neese2009c,Izsak2011,Izsak2012,Dutta2016,Izsak2013,Kossmann2010,Kossmann2009} and in the approximation of many-electron integrals in explicitly correlated F12 methods.\cite{TenNo:2004dma}
    
    Computing $g_{ab,cd}$ using numerical quadrature involves replacing the integration over a single electron, for example electron 1, with a sum over a set of quadrature points:
    \begin{align}
      \label{eq:gabcd-ps-rs}
      g_{ab,cd} \overset{\text{PS}}{\approx} & \sum_{g} w_{g} \phi_a^*({\bf r}_g) \phi_b({\bf r}_g) \int g({\bf r}_g,{\bf r}_2) \phi_c^*({\bf r}_2) \phi_d({\bf r}_2) \, d{\bf r}_2 ;
    \end{align}
    Introducing
    \begin{align}
    X_{a,g} \equiv & \sqrt{w_{g}} \phi_a({\bf r}_g), \\
    Y_{g,cd} \equiv & \int g({\bf r}_g,{\bf r}_2) \phi_c^*({\bf r}_2) \phi_d({\bf r}_2) \, d{\bf r}_2,
    \end{align}
    leads to the {\em algebraic} form of the PS approximation,
    \begin{align}
    \label{eq:gabcd-ps-alg}
      g_{ab,cd} \overset{\text{PS}}{\approx} & \sum_g X_{a,g}^* X_{b,g} Y_{g,cd},
    \end{align}
    which makes the connection to CP factorization obvious; note that the summation over grid points $g$ corresponds to the 3-way {\em hyperedge} in the diagrammatic representation of \cref{eq:gabcd-ps-alg} in \cref{fig:g_cp_ps}. In practice, an accurate implementation of the PS approximation is sensitive to choice of grid and requires various measures to reduce the error.\cite{Neese2009c,Parrish2012,Greeley1994,Izsak2011,Izsak2013}
    However, the algebraic form of the PS approximation can be viewed as an abstract tensor network approximation of $g_{ab,cd}$,
    with factors $X$ and $Y$ defined not by the particular choice of real-space quadrature in \eqref{eq:gabcd-ps-rs},
    but by arbitrary fitness conditions.
    
    Inserting a quadrature once for {\em every} particle leads to, what Martinez and co-workers termed, the tensor hypercontraction\footnote{The term ``hypercontraction'' presumably refers to the appearance of {\em hyperedges} in the diagrammatic representation of CP-like tensor networks, e.g., \Cref{fig:g_cp_ps}.} (THC) approximation\cite{Hohenstein2013,Schutski2017,Parrish:2013iz,Hohenstein2012,Parrish2012,Hohenstein2012a,Lee2019,Parrish2014,Shenvi2013} of $g_{ab,cd}$,
    \begin{align}
      \label{eq:gabcd-thc-rs}
    g_{ab,cd} \overset{\text{THC}}{\approx} & \sum_{g_1,g_2} w_{g_1} w_{g_2} \phi_a^*({\bf r}_{g_1}) \phi_b({\bf r}_{g_1}) g({\bf r}_{g_1},{\bf r}_{g_2}) \phi_c^*({\bf r}_{g_2}) \phi_d({\bf r}_{g_2}),
    \end{align}
    and its algebraic form:
    \begin{align}
    \label{eq:gabcd-thc-alg}
    g_{ab,cd} \overset{\text{THC}}{\approx} & \sum_{g_1} \sum_{g_2} X_{a,g_1}^* X_{b,g_1} Y_{g_1,g_2} X_{c,g_2}^* X_{d,g_2}.
    \end{align}
    The diagrammatic representation of \cref{eq:gabcd-thc-alg}, shown in \cref{fig:g_cp_df}, includes {\em two} 3-way hyperedges.
    Clearly, the same idea can be applied to a matrix element of any (local) $n$-body operator.\cite{Parrish:2013iz}
    THC approximation was originally exploited in the algebraic form, using algebraic CP decomposition of 3-center overlap integrals in the context of (non-robust) overlap-metric DF to define factors $X$ and $Y$ in \cref{eq:gabcd-thc-alg} (``PF-THC'').\cite{Hohenstein2012} It was subsequently formulated using real-space quadrature to define factors $X$ and least-squares fitting to determine factor $Y$ in \cref{eq:gabcd-thc-alg} (``LS-THC'')\cite{Parrish2012,Parrish2014,Lee2019}. What these approaches have in common with each other and with other related factorizations\cite{Hummel2017}
     is use of the tensor network topology of \cref{eq:gabcd-thc-alg}; how the factors are determined can differ widely between the methods.

    Although our focus in this manuscript is on the 3-way CP factorization (CP3) we should also note that the direct 4-way algebraic CP factorization of Coulomb integrals (CP4) has been employed by Benedikt and co-workers.\cite{Benedikt2013,Benedikt2013a,Benedikt2011} Related 4-way factorizations of Coulomb integrals has been considered by Peng and Kowalski, who proposed to compress the Cholesky factors of the Coulomb tensor by the SVD; the use of factorized integrals has been explored in the CC method.\cite{Peng:2017iy} More recently, Motta and co-workers employed a similar multi-step factorization to reduce the cost of auxiliary-field Quantum Monte Carlo methods.\cite{Motta2019}  
    The similarity of these factorizations to the 4-way CP decomposition due to the appearance of the 4-way hyperedge, whereas all of the factorizations considered in this work are limited to 3-way hyperedges only.
    
    To introduce the main result of our work consider how to best introduce the CP3 approximation (\cref{eq:B_cp3_symm}) for the symmetric (CD/DF-like) factorization in \cref{eq:df_approx1}. Using CP3 {\em once} produces a PS-like factorization, to which we will refer as CP-PS:
    \begin{align}
    \label{eq:g_cp_ps}
    g_{ab,cd} \overset{\text{CP-PS}}{\approx} \sum_X \sum_r^R \beta_{a,r} \beta_{b,r} \gamma_{X,r} B_{cd,X} = \sum_r^R \beta_{a,r} \beta_{b,r} (\gamma B)_{cd,r} ,
    \end{align}
    where we introduced
    \begin{align}
    \label{eq:gammaB}
    (\gamma B)_{cd,r} \equiv \sum_X \gamma_{X,r} B_{cd,X};
    \end{align}
    compare \cref{eq:g_cp_ps} to \cref{eq:gabcd-ps-alg} to recognize the connection to the algebraic PS factorization.
    Using CP3 {\em twice} produces a THC-like factorization, to which we will refer as CP-DF:
    \begin{align}
    \label{eq:g_cp_df}
    g_{ab,cd} \overset{\text{CP-DF}}{\approx} \sum_X \sum_r^R \beta_{a,r} \beta_{b,r} \gamma_{X,r} \sum_{r'}^R \beta_{c,r'} \beta_{d,r'} \gamma_{X,r'} = \sum_r^R \beta_{a,r} \beta_{b,r} \sum_{r'}^R \beta_{c,r'} \beta_{d,r'} (\gamma\gamma^\top)_{r,r'} ,
    \end{align}
    where we introduced $(\gamma\gamma^\top)_{r,r'} \equiv \sum_X \gamma_{X,r} \gamma_{X,r'}$; compare \cref{eq:g_cp_df} to \cref{eq:gabcd-thc-alg} to recognize the connection to the algebraic THC factorization.

    Clearly, both CP-PS and CP-DF approximations are linear in the error introduced by the CP3 approximation (\cref{eq:B_cp3_symm}). As discussed in
    \Cref{sec:robusttnapp}, it is possible to eliminate the linear error using the {\em robust} form of \text{CP-DF}, to which we will refer as rCP-DF:
    \begin{align}
    \label{eq:g_rcp_df}
    g_{ab,cd} \overset{\text{rCP-DF}}{\approx} 2 g_{ab,cd}^{\text{CP-PS}} - g_{ab,cd}^{\text{CP-DF}} =
    \sum_r^R \beta_{a,r} \beta_{b,r} \left( 2 (\gamma B)_{cd,r} - \sum_{r'}^R \beta_{c,r'} \beta_{d,r'} (\gamma \gamma^\top)_{r,r'} \right).
    \end{align}
    Although the rCP-DF approximant has a higher computational cost, than either CP-PS or CP-DF, computing the PPL diagram with the rCP-DF approximation has the same complexity (\bigO{N^5}) as the aforementioned approaches. However, the systematic error cancellation unique to rCP-DF should, at equal CP rank, result in significantly smaller errors than either CP-PS or CP-DF and thus should be computationally superior to these simpler alternatives.
    
  \subsection{Application to the particle-particle ladder diagram}
    Our primary objective is to reduce
    the computational cost of the particle-particle ladder (PPL) diagram in CC and other many-body methods. It is well known that both PS\cite{Martinez1993, Martinez1995} and THC factorizations\cite{Parrish2014,Hummel2017} can reduce the computational complexity of the PPL term in the canonical MO basis from \bigO{N^6} to \bigO{N^5}, hence the same should be possible for the PPL term in the rCP-DF approximation. Indeed, plugging in \cref{eq:g_cp_ps} into
    the spin-free PPL expression (permutational symmetry is ignored for simplicity) yields:
    \begin{align}
    \label{eq:ppl_cp_ps}
    \sum_{bd} g_{ab,cd} t_{bdij} \overset{\text{CP-PS}}{\approx} \text{PPL}^\text{CP-PS} \equiv \sum_r^R \beta_{a,r} \left( \sum_d (\gamma B)_{cd,r} \left( \sum_b \beta_{b,r} t_{bdij} \right) \right).
    \end{align}
    The order of evaluation which minimizes the operation count is shown by parentheses, with the result of each binary tensor product
    stored in an intermediate tensor. The inner-most product, $\sum_b \beta_{b,r} t_{bdij} \to (I_1)_{rdij}$, is {\em covariant} (i.e., it is a pure tensor contraction) and has an operation cost of $2 o^2 u^2 R$, where $o$ and $u$ are the numbers of occupied and unoccupied MOs,
    respectively, and $R$ is the CP rank. The second product is of general type (i.e., it cannot be mapped to a single matrix multiplication), and has the same cost as the first product. The last product is a pure contraction and has the same cost as the other 2 contractions. The total operation count of the CP-PS approximated PPL is thus $6 o^2 u^2 R$ vs the $2 o^2 u^4$ cost of the naive approach; note that precomputing the $(\gamma B)$ intermediate (\cref{eq:gammaB}) is done once, outside of the CCSD solver loop, and has the negligible cost ($2u^2 X R$, where $X$ is the size of the DF fitting basis). We can expect computational savings from the use of CP-PS when $R < u^2/3$.\footnote{Note that the CP-PS approximation breaks particle equivalence symmetry and therefore, in practice, the result must be symmetrized with respect to the transpose of $ia$ and $jc$ index pairs.}
    
    The PPL term can be similarly reformulated with the \bigO{N^5} cost using the CP-DF approximation. One approach, utilized by Parrish et al.\cite{Parrish2014} and Hummel et al.,\cite{Hummel2017}, uses the CP-PS route (\cref{eq:ppl_cp_ps}) by recomputing the appropriate intermediates:
    \begin{align}
    \label{eq:ppl_cp_df1}
    \sum_{bd} g_{ab,cd} t_{bdij} \overset{\text{CP-DF}}{\approx} \text{PPL}^\text{CP-DF} \equiv \sum_r^R \beta_{a,r} \left( \sum_d (\gamma \hat{B})_{cd,r} \left( \sum_b \beta_{b,r} t_{bdij} \right) \right),
    \end{align}
    where $(\gamma \hat{B})_{cd,r}$ is the CP-factorized intermediate $(\gamma B)_{cd,r}$, obtained by inserting \cref{eq:B_cp3_symm} into \cref{eq:gammaB}\footnote{N.B. if $5X > 3R$, \cref{eq:gammahatB} can be reordered to compute $(\gamma \hat{B})$ more efficiently}:
    \begin{align}
    \label{eq:gammahatB}
    (\gamma \hat{B})_{cd,r} \equiv & \sum_X \gamma_{X,r} \left( \sum_{r'}^R \beta_{a,r'} \beta_{b,r'} \gamma_{X,r'} \right).
    \end{align}
    The operation count of this route is $6 o^2 u^2 R$, hence the crossover relative to the naive PPL evaluation occurs at the same CP rank as in the CP-PS route.
    
    Another CP-DF route, utilized by Hummel et al.\cite{Hummel2017} and Mardirossian et al.\cite{Mardirossian:2018fk} introduces order-4 tensors with 2 CP indices:
    \begin{align}
    \label{eq:ppl_cp_df2}    
    \sum_{bd} g_{ab,cd} t_{bdij} \overset{\text{CP-DF}}{\approx} \text{PPL}^\text{CP-DF} \equiv \sum_r^R \beta_{a,r} \sum_{r'}^R \left( \beta_{c,r'} \left( (\gamma \gamma^\top)_{r,r'} \left( \sum_b \beta_{b,r} \left( \sum_d \beta_{d,r'} t_{bdij} \right) \right) \right) \right).
    \end{align}
    Compared to 3 tensor products in the CP-PS approach, the CP-DF route has 5 products, with all but the third product of $(\gamma \gamma^\top)$ being pure contractions.
    The operation count is $4 o^2 u^2 R + 4 o^2 u R^2 + o^2 R^2$; since in practice $R \gg u$, the cost is expected to be dominated by the $4 o^2 u R^2$ contribution.
    
    To reduce the operation count, relative to the conventional PPL, the route outlined above requires $R < \sqrt{u^3/2} = u^{3/2}/\sqrt{2}$ (compared to $R < u^2 / 3$ requirement of the CP-PS-based route). Clearly, the cost crossover occurs earlier in the CP-PS-based route. Furthermore, the low arithmetic intensity of the element-wise (Hadamard-like) third product in \cref{eq:ppl_cp_df2} lowers the computational efficiency of this approach. For these reasons, throughout our work we used the CP-PS-based approach, \cref{eq:ppl_cp_df1}, to implement CP-DF PPL.
    
    Clearly, the PPL term can be therefore approximated via rCP-DF with the \bigO{N^5} cost by naively combining the CP-PS and CP-DF approximations:
    \begin{align}
    \label{eq:ppl_rcp_df0}
    \sum_{bd} g_{ab,cd} t_{bdij} \overset{\text{rCP-DF}}{\approx} 2 \times \text{PPL}^\text{CP-PS} - \text{PPL}^\text{CP-DF}.
    \end{align}
    Plugging \cref{eq:ppl_cp_ps} and \cref{eq:ppl_cp_df1} into \cref{eq:ppl_rcp_df0} and refactoring leads to the following evaluation scheme with optimal operation count:
    \begin{align}
    \label{eq:ppl_rcp_df1}
    \sum_{bd} g_{ab,cd} t_{bdij} \overset{\text{rCP-DF}}{\approx} \text{PPL}^\text{rCP-DF} \equiv \sum_r^R \beta_{a,r} \left( \sum_d (\gamma \tilde{B})_{cd,r} \left( \sum_b \beta_{b,r} t_{bdij} \right) \right),
    \end{align}
    in which we introduced
    \begin{align}
    (\gamma \tilde{B})_{cd,r} \equiv 2 (\gamma B)_{cd,r} - (\gamma \hat{B})_{cd,r}
    \end{align}
    The total operation count of the rCP-DF PPL approximation is $6 o^2 u^2 R$, which is identical to that of the CP-PS and CP-DF PPL approximations. Thus, rCP-DF is the preferred 3-way CP approach in the context of the PPL evaluation.
    
\section{Computational Details} \label{sec:comp_det}

  CP approximations for order-3 tensors were computed using the standard alternating least squares (ALS) method.\cite{Kroonenberg:1980il,Beylkin2002} Although ALS can be slow to converge and the quality of the solution can strongly depend on the initial guess,\cite{Uschmajew2012} we found that our solver converged robustly with an initial guess of vectors generated using quasi-random numbers taken from the uniform distribution on [-1,1]. No consistent benefit was found from an initial guess scheme which generated factor matrices using the higher-order SVD (HOSVD)\cite{Kolda2008} padded with random vectors (where random vectors were generated as just described). Furthermore, no discernible benefit was found from the use of a regularized ALS (RALS) solver.\cite{Navasca2008} The use of non-linear and gradient-based solvers\cite{Acar2009,Sorber2013} as an alternative to ALS will be investigated in future work. 
  
  Assessment of the CP-based Coulomb tensor factorizations utilized the full S66 benchmark set of weakly-bound complexes\cite{Rezac2011a} as well as a 12-system representative set of 12 complexes (S66/12)\footnote{1 water \dots water, 2 water \dots \ce{MeOH}, 3 water \dots \ce{MeNH2}, 4 \ce{MeNH2} \dots \ce{MeOH}, 5 benzene \dots benzene ($\pi$-$\pi$), 6 pyridine \dots pyridine ($\pi$-$\pi$), 7 uracil \dots uracil ($\pi$-$\pi$), 8 pentane \dots pentane, 9 benzene \dots benzene (TS), 10 benzene \dots ethyne (CH-$\pi$), 11 ethyne \dots water (CH-O), 12 \ce{MeNH2} \dots pyridine}; some computations utilized a smaller 7-system subset of S66/12 (systems 1-4 and 10-12; dubbed S66/7). The S66 geometries were taken from the Benchmark Energy and Geometry Database (BEGDB).\cite{Rezac2008} Additional assessments utilized the HJO12 set of isogyric reaction energies,\cite{Helgaker:2000db,Zhang2012}
  the 8 low-lying conformers of (\ce{H2O})$_{6}$\cite{Bates2011} and a conformer of (\ce{H2O})$_{20}$.\cite{Jorgensen1983,Wales1998} 
  All of the above computations utilized the cc-pVDZ-F12 (abbreviated as DZ-F12) orbital basis set (OBS)\cite{Peterson2008}. The 2-electron interaction tensors were approximated using standard Coulomb-metric density fitting using the aug-cc-pVDZ-RI (abbreviated as aVDZ-RI) density fitting basis set (DFBS).\cite{Weigend2002a}
  Assessment of the basis set variation in the performance of rCP-DF used the following additional OBS/DFBS pairs: the aug-cc-pVDZ\cite{Dunning1989,Kendall1992} (aVDZ) OBS paired with the aVDZ-RI DFBS, the aug-cc-pVTZ (aVTZ) OBS\cite{Dunning1989,Kendall1992} paired with the aug-cc-pVTZ-RI\cite{Weigend2002a} (aVTZ-RI) DFBS, and  the cc-pVTZ-F12\cite{Peterson2008} (TZ-F12) OBS paired with the aVTZ-RI DFBS. The CP approximations of Coulomb integral tensors was utilized in {\em only} the PPL diagram of CCSD . Only valence electrons were correlated in all CCSD computations. 
  
 All computations were run on the Virginia Tech Advanced Research Computing's Cascades cluster which utilizes standard nodes that contain 2 Intel Xeon E5-2683 v4 CPUs, and high-memory nodes, each with 4 Intel Xeon E7-8867 v4 CPUs. Only the (\ce{H2O})$_{20}$ computations utilized Cascades high-memory nodes. In the following section, speedup is determined as
  \begin{equation} \label{eqn:speedup}
    \text{speedup} = \frac{t_{\text{DF-CCSD}}} {t_{\text{CP-PPL-DF-CCSD}} + t_{\text{CP-ALS}}}
  \end{equation}
  where $t_{\text{DF-CCSD}}$ and $t_{\text{CP-PPL-DF-CCSD}}$ are the total time it takes to compute the CCSD correlation energy with either the DF or CP approximation applied to the PPL diagram and $t_{\text{CP-ALS}}$ is the time it takes to compute the CP decomposition using the ALS method.

  The CP-ALS decomposition was implemented in C++ in the open-source Basic Tensor Algebra Subroutines (BTAS) library.\cite{BTAS}
  The CP-DF, CP-PS and rCP-DF approximations are implemented in a developmental version of the Massively Parallel Quantum Chemistry (MPQC) package.\cite{Peng2020}

\section{Results} \label{sec:results}

  The discussion of computational experiments is organized as follows. In \Cref{sec:results_g} we examine how the errors in the matrix elements of the Coulomb operator converge with respect to the CP rank. It turns out that the use of CP in the CP-PS and CP-DF approximations results in 2 types of errors: due to suboptimal factors in the tensor network and due to the deficient CP rank; the use of the robust approximation greatly reduces both types of errors. In \Cref{sec:results_ccsd_errors,sec:results_ccsd_cost} we discuss the error in the CCSD energies introduced by and the cost reduction of the CP approximation of the PPL diagram, respectively. Note, to standardize CP rank across systems, we report the CP rank in the units of $X$ (the size of the density fitting basis), which grows proportionally to $n$. 
  
  \subsection{Errors in Coulomb matrix elements: effects of CP factor optimality, CP rank, and robustness \label{sec:results_g}}
  
\begin{figure}[ht]
      \includegraphics[width=0.9\textwidth]{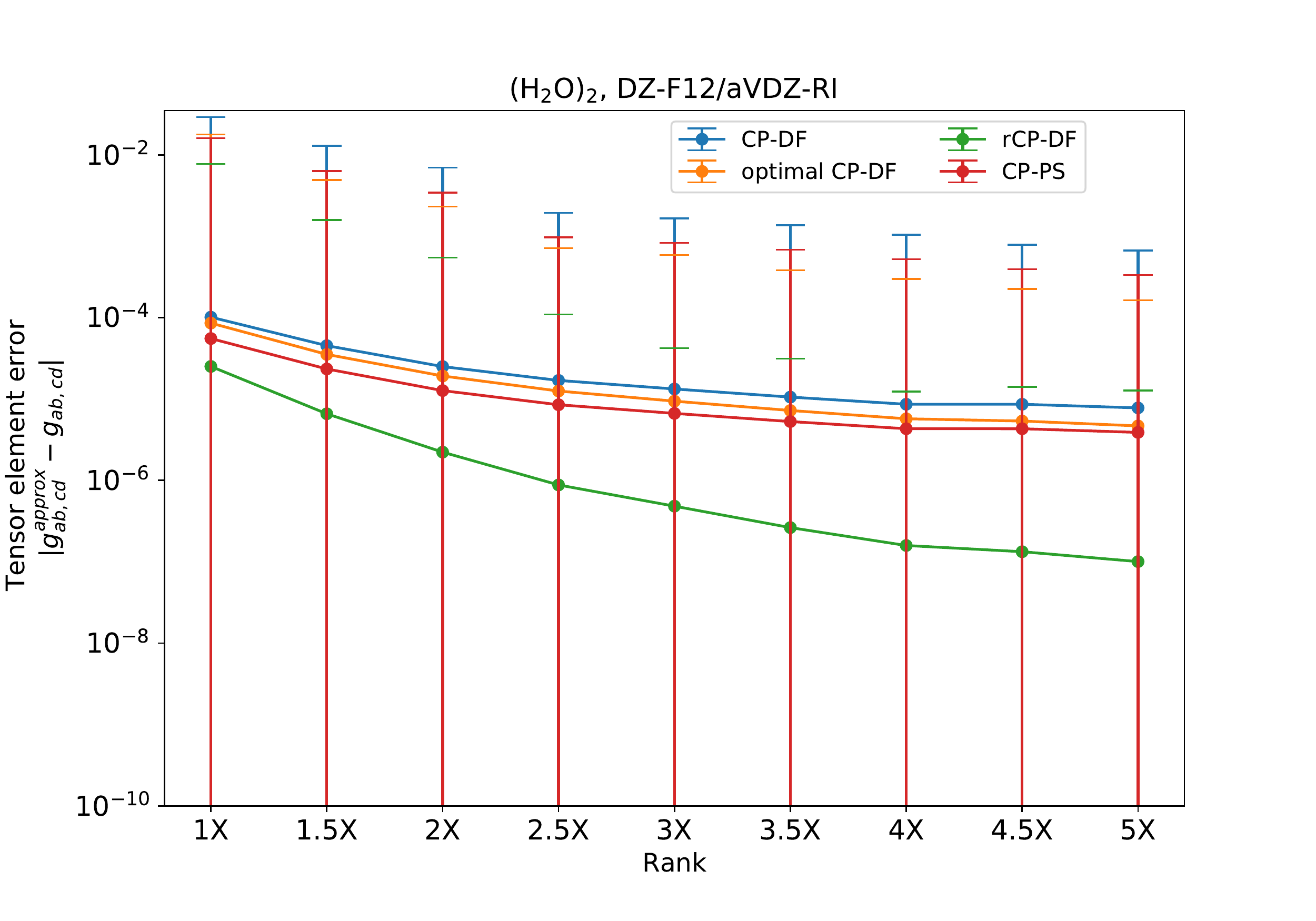}
      \caption{Absolute errors in matrix elements of $g_{ab,cd}$ for a water dimer with S66 configuration approximated by the CP-PS, CP-DF, and rCP-DF factorizations obtained with ALS precision of $\epsilon = 10^{-3}$. The error bars denote the max/min unsigned errors.}
      \label{fig:element_errors_water_dimer}
  \end{figure}

  The most direct way to assess a particular factorization of the Coulomb interaction tensor is to examine the matrix elements themselves. Since the data varies little between systems, \cref{fig:element_errors_water_dimer} shows the absolute errors of the matrix elements of $g_{ab,cd}$ for a particular system, namely, the water dimer at the S66 geometry. The first observation is that both the average (solid circles) and the maximum (horizontal line) errors decrease in the CP-DF>CP-PS>rCP-DF series, with the CP-DF and CP-PS errors decaying with the CP rank at a similar rate, and much slower than the rCP-DF errors. This observation is easy to explain. Using the matrix notation introduced in \cref{eq:df_approx2}, it is clear that the leading-order error of the CP-DF factorization should be roughly twice the error of CP-PS:
  \begin{align}
  {\bf g}^\text{DF} - {\bf g}^\text{CP-PS} = & \mathbf{B} \mathbf{B}^\top - \frac{1}{2} \left( \hat{\mathbf{B}} \mathbf{B}^\top + \mathbf{B} \hat{\mathbf{B}}^\top \right)= \frac{1}{2} \left( \pmb{\delta} \mathbf{B}^\top + \mathbf{B} \pmb{\delta}^\top \right) , \\
  {\bf g}^\text{DF} - {\bf g}^\text{CP-DF} = & \mathbf{B} \mathbf{B}^\top - \hat{\mathbf{B}} \hat{\mathbf{B}}^\top = \pmb{\delta} \hat{\mathbf{B}}^\top + \hat{\mathbf{B}} \pmb{\delta}^\top + \pmb{\delta} \pmb{\delta}^\top = 2 \left({\bf g}^\text{DF} - {\bf g}^\text{CP-PS} \right) + \pmb{\delta} \pmb{\delta}^\top,
  \end{align}
  where $\hat{\mathbf{B}}$ is the matricized form of the CP approximant in \cref{eq:B_cp3_symm}, and
  \begin{align}
  \label{eq:deltamat}
  \pmb{\delta} \equiv \mathbf{B} - \hat{\mathbf{B}}
  \end{align}
  is the CP error tensor. Clearly, as the CP rank increases, the CP error $\pmb{\delta}$ decreases but the CP-PS / CP-DF ratio of errors stays approximately 2. Since the rCP-DF is quadratic in $\pmb{\delta}$, the rCP-DF error should decay with the CP rank faster than either that of CP-PS or CP-DF. The improvement of rCP-DF over CP-DF is approximately one order of magnitude for $R=1.5 X$, and approaches 2 orders of magnitude for $R=5X$.  
  
  It is instructive to wonder whether it is possible to improve CP-PS and CP-DF approximations solely by relaxing the factors in the respective tensor networks approximating $g_{ab,cd}$. Indeed, it is important to recognize that CP-PS and CP-DF approximations utilize CP factorization of $\mathbf{B}$ that is optimal (in the least-squares sense) for representing $\mathbf{B}$, not $\mathbf{g}$. It is therefore possible to optimize the factors in the tensor networks approximation of $\mathbf{g}$ directly. Partial relaxation of the factors in the CP-PS and CP-DF networks to minimize the error in $\mathbf{g}$ was already employed in some real-space-based THC developments by Parrish et al.,\cite{Parrish2012,Parrish2014} and full relaxation of the CP-DF network cost was implemented by Schutski et al.\cite{Schutski2017} (e.g., see the discussion of their THC-ALS-RI solver). To investigate whether the suboptimality of the CP-DF network using the $\mathbf{B}$-optimized factors is significant we implemented an ALS solver that minimizes the CP-DF error in $\mathbf{g}^\text{DF}$;\footnote{See the Supporting Information for the detailed algorithm description.} the operation complexity of such solver is identical to the \bigO{N^4} complexity of the ALS solver for the CP decomposition of $\mathbf{B}$, albeit the prefactor is somewhat larger. Only few iterations are needed to relax the CP-DF network fully with respect to $\mathbf{g}$ if we use, as the initial guess, the factors obtained by CP3 decomposing $\mathbf{B}$.
  
  As the data in \cref{sec:results_ccsd_errors,sec:results_ccsd_cost} indicates, the tensor element errors obtained with the $\mathbf{g}$-optimized CP-DF network are moderately smaller than the errors of the reference CP-DF network, but still exceed the CP-PS errors and they are not competitive with the errors in the zero-cost robust CP-DF approximant. This observation suggests that the dominant source of error in the CP-DF (and CP-PS) approximants is the deficiency of the CP rank. The robust approximation is clearly able to greatly reduce both sources of error, due to the suboptimality (with respect to $\mathbf{g}^\text{DF}$) of the factors in the CP-DF network and due to the deficient CP rank.
  
  \subsection{Errors in the CCSD energies vs. the CP approximation parameters \label{sec:results_ccsd_errors}}

    The error of the CP approximation is determined by the CP rank, $R$, and by the precision, $\epsilon$, of the inexact CP solver (in our case, ALS); as already mentioned we found negligible dependence of the ALS solution on the initial random guess. The ALS precision in this work is estimated by the difference between the current and previous iteration's decomposition ``fit'' $\Delta$ defined for \cref{eq:B_cp3_symm} as
      \begin{align}
        \Delta \equiv 1.0 - \frac{\| B_{ab,X} - \sum_r^R \beta_{a,r} \beta_{b,r} \gamma_{X,r} \| } {\| B_{ab,X}\|} = 1.0 - \frac{\|\pmb{\delta}\|}{\|B_{ab,X}\|}
      \end{align}
    where $\pmb{\delta}$ is the CP error tensor as defined in \cref{eq:deltamat}. Clearly, because $\epsilon$ depends on the {\em change} in the loss function, smaller values for $\epsilon$ do not necessarily lead to a smaller CP error. Thus, we first assessed how the error in $E_{\text{CCSD}}$ due to the CP approximation depends on $\epsilon$ for a range of fixed CP ranks, $R$.

     \subsubsection{Variation of the CP error with the ALS solver precision \label{sec:results_ccsd_errors_eps}}
     
\begin{figure}[ht]
        \begin{subfigure}[t]{0.45\textwidth}
          \includegraphics[width=0.9\textwidth]{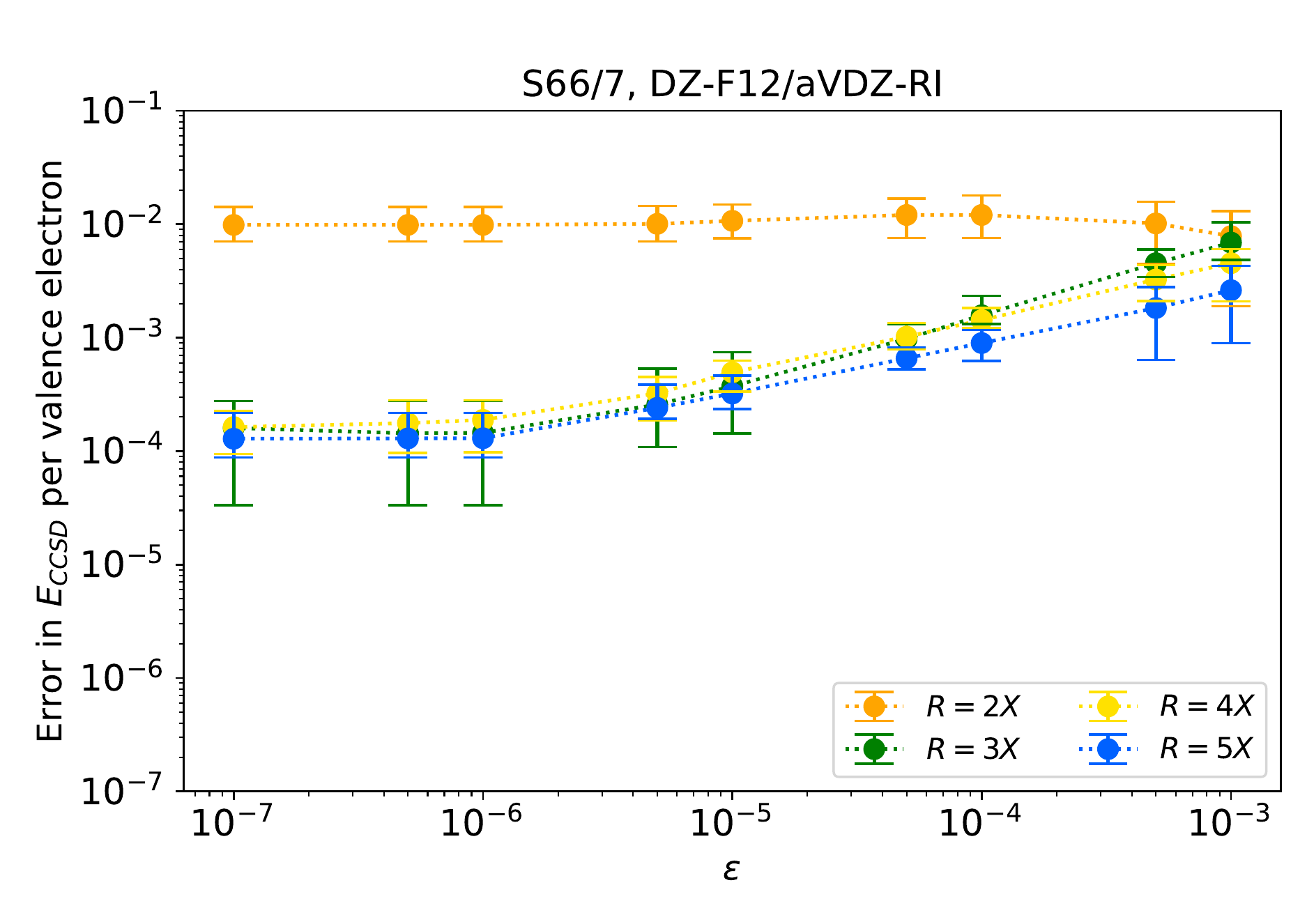}
          \caption{}
          \label{fig:ep_ps}
        \end{subfigure} \hfill
        \begin{subfigure}[t]{0.45\textwidth}
          \includegraphics[width=0.9\textwidth]{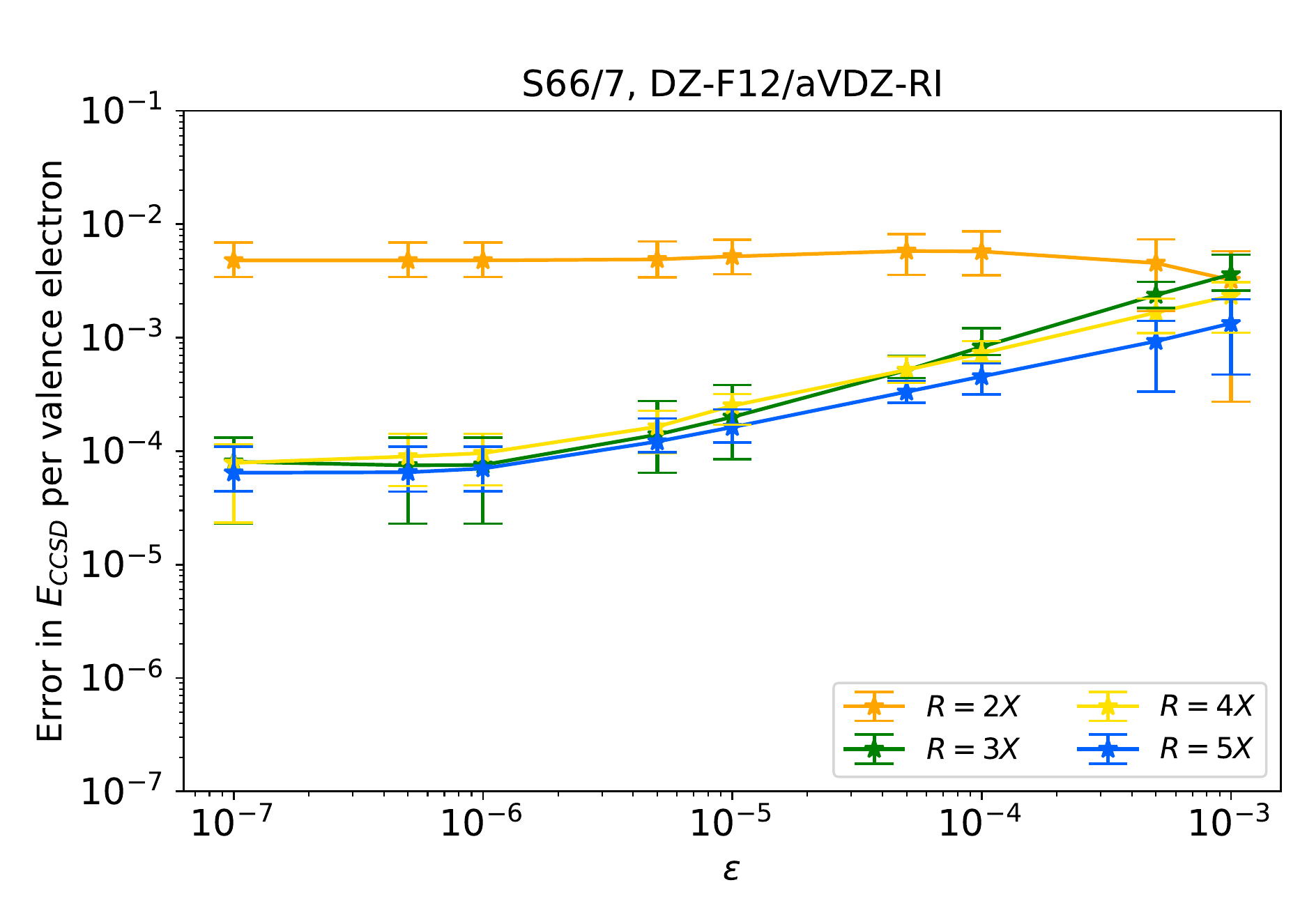}
          \caption{}
          \label{fig:ep_df}
        \end{subfigure}
        \begin{subfigure}{0.45\textwidth}
          \includegraphics[width=0.9\textwidth]{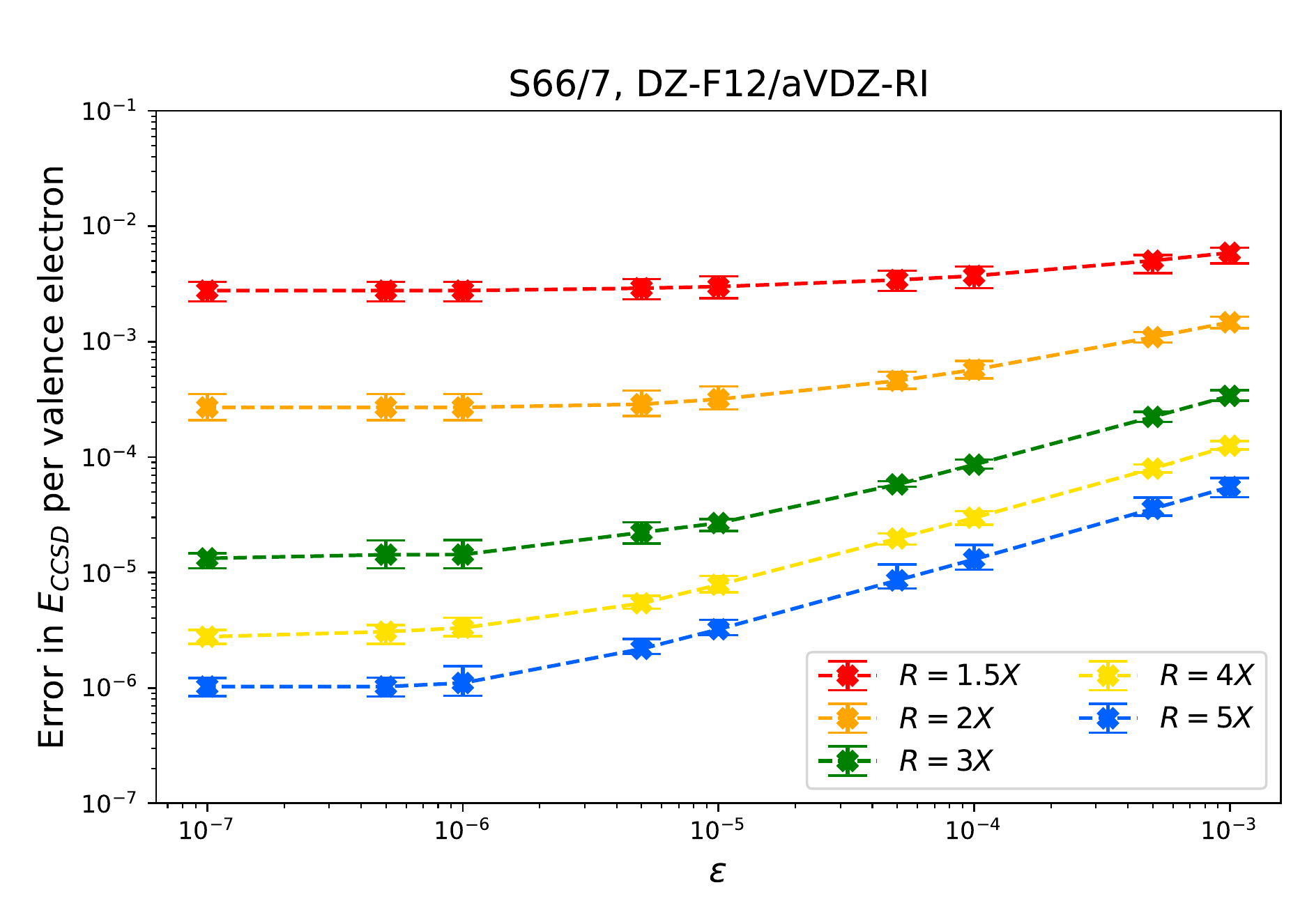}
          \caption{}
          \label{fig:ep_rcp}
        \end{subfigure}
        \caption{Mean unsigned errors in the per-electron CCSD correlation energies (kcal/mol) of molecules in the S66/12 dataset, relative to canonical CCSD, induced by the (a) CP-DF, (b) CP-PS or (c) rCP-DF approximations to PPL vs the ALS precision ($\epsilon$). The error bars denote the max/min unsigned errors.}
        \label{fig:ep_test}
      \end{figure}

         \cref{fig:ep_test} report the relationship between $\epsilon$ and the CP error in the valence CCSD correlation energy per electron for the
        S66/7 test set for CP ranks in the $X \leq R \leq 5X$ range.    
        For low CP ranks ($R \leq 2 X$) the error varies little with $\epsilon$. As CP rank increases progressively smaller values of $\epsilon$ are required to obtain sufficiently converged ALS solutions. However, the effect of $\epsilon$ on the CCSD energy is significantly weaker than that of the CP rank $R$.
    
     \subsubsection{Variation of the CP error with the CP rank \label{sec:results_ccsd_errors_rank}}
    
        \cref{fig:ep_test} indicated that increasing the CP rank $R$ reduced the error in the CCSD energy monotonically. These figures also gave the first evidence of performance advantage of rCP-DF over CP-DF and CP-PS. At $R = 1.5X$ (the red line in \cref{fig:ep_rcp}), rCP-DF is more accurate than both CP-DF and CP-PS with $R = 2X$ (the orange line in \cref{fig:ep_df,fig:ep_ps}). Furthermore, the error in CCSD energy is reduced at a fast rate, with respect to CP rank, for rCP-DF which corroborates our discussion in \cref{sec:results_g}. For each $R$ and at converged $\epsilon$, the rCP-DF approximation introduces error which is at least an order of magnitude smaller than the error introduced by either CP-DF or CP-PS.
    
\begin{figure} 
     \begin{subfigure}[t]{0.45\textwidth}
      \includegraphics[width=0.9\linewidth]{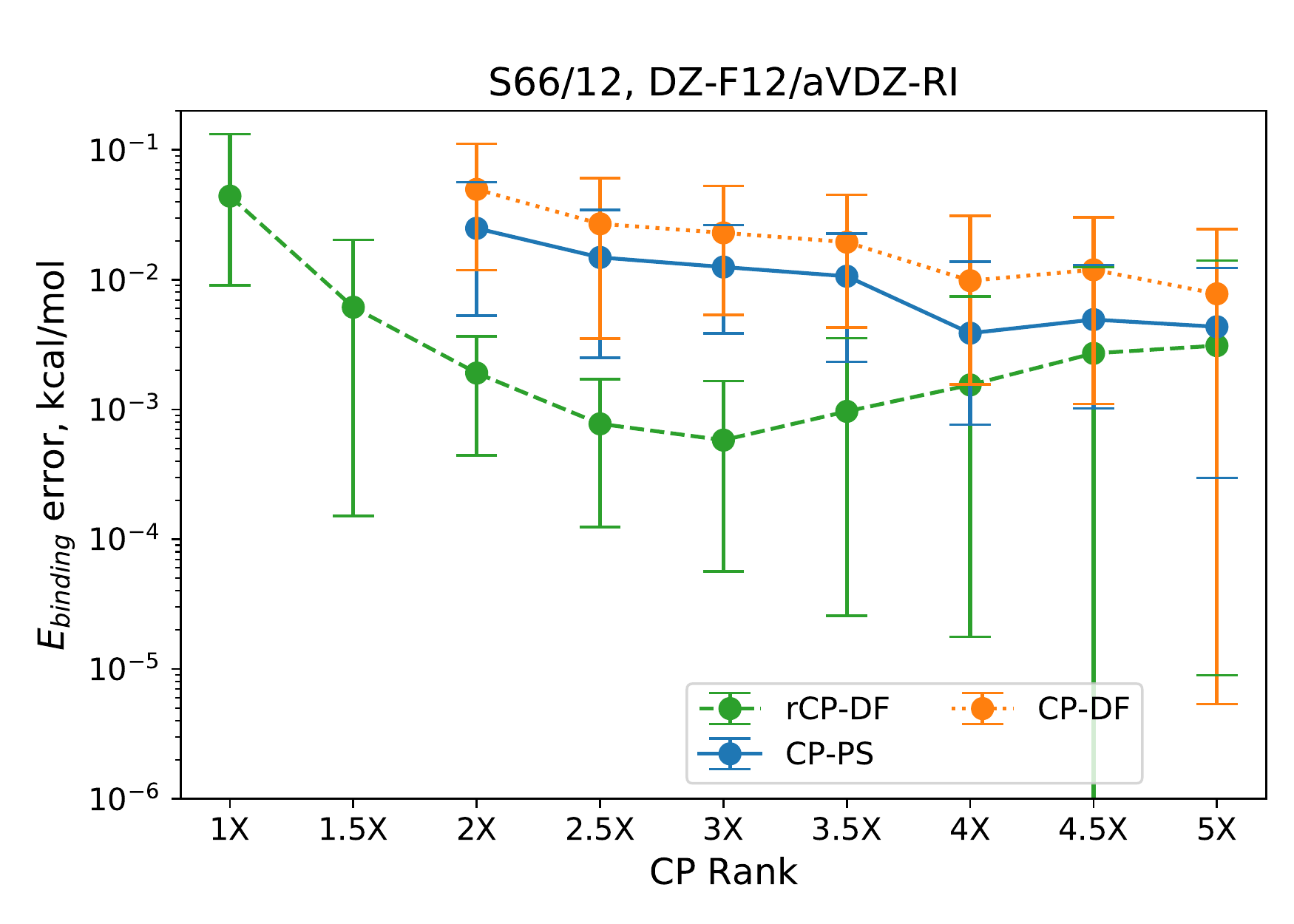}
      \caption{}
      \label{fig:BE-us-error-1e3}
     \end{subfigure} \hfill
     \begin{subfigure}[t]{0.45\textwidth}
      \includegraphics[width=0.9\linewidth]{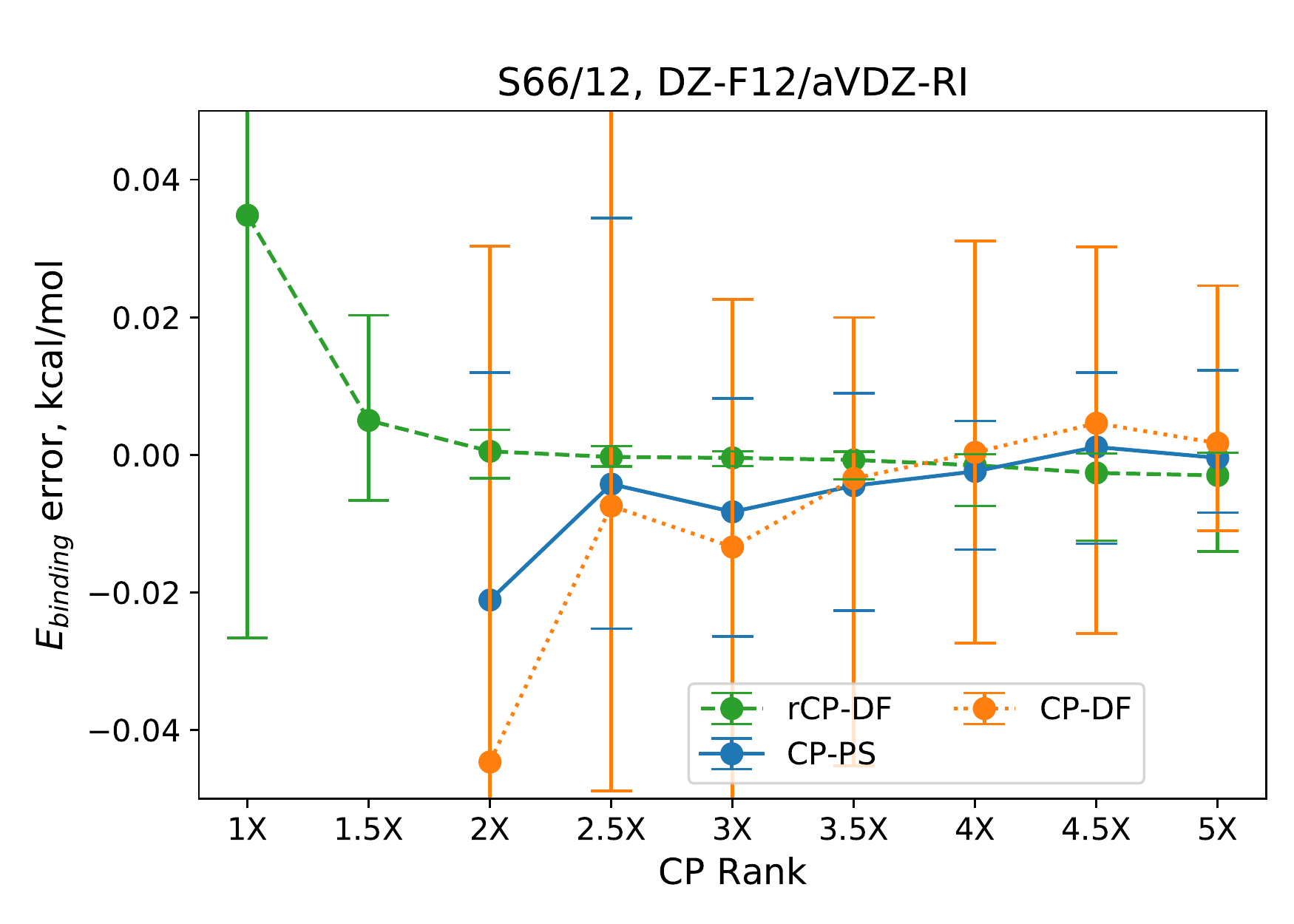}
      \caption{}
      \label{fig:BE-s-error-1e3}
     \end{subfigure}
     \caption{
     Mean unsigned (a) and signed (b) errors, respectively, in the CCSD binding energies (kcal/mol) of the S66/12 dataset, relative to canonical CCSD, induced by the CP-DF, CP-PS or rCP-DF approximations to PPL vs CP rank $R$ (in units of the fitting basis, $X$). ALS precision fixed at $\epsilon = 10^{-3}$. The error bars denote the max/min errors.}
     \label{fig:BE}
  \end{figure}

\begin{figure}
    \begin{subfigure}[t]{0.45\textwidth}
      \includegraphics[width=0.9\linewidth]{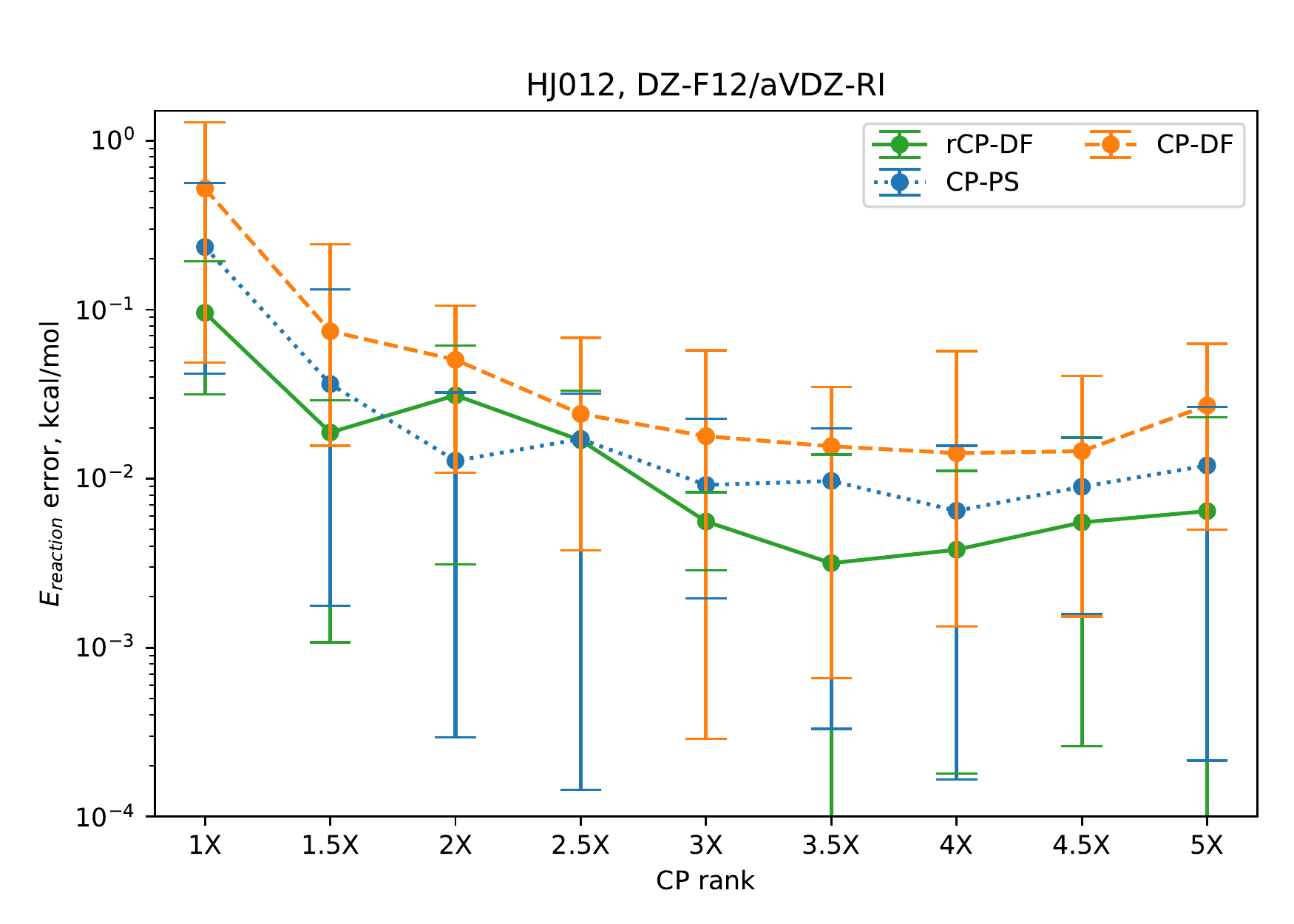}
      \caption{}
      \label{fig:RE-error-1e3}
    \end{subfigure} \hfill
    \begin{subfigure}[t]{0.45\textwidth}
      \includegraphics[width=0.9\linewidth]{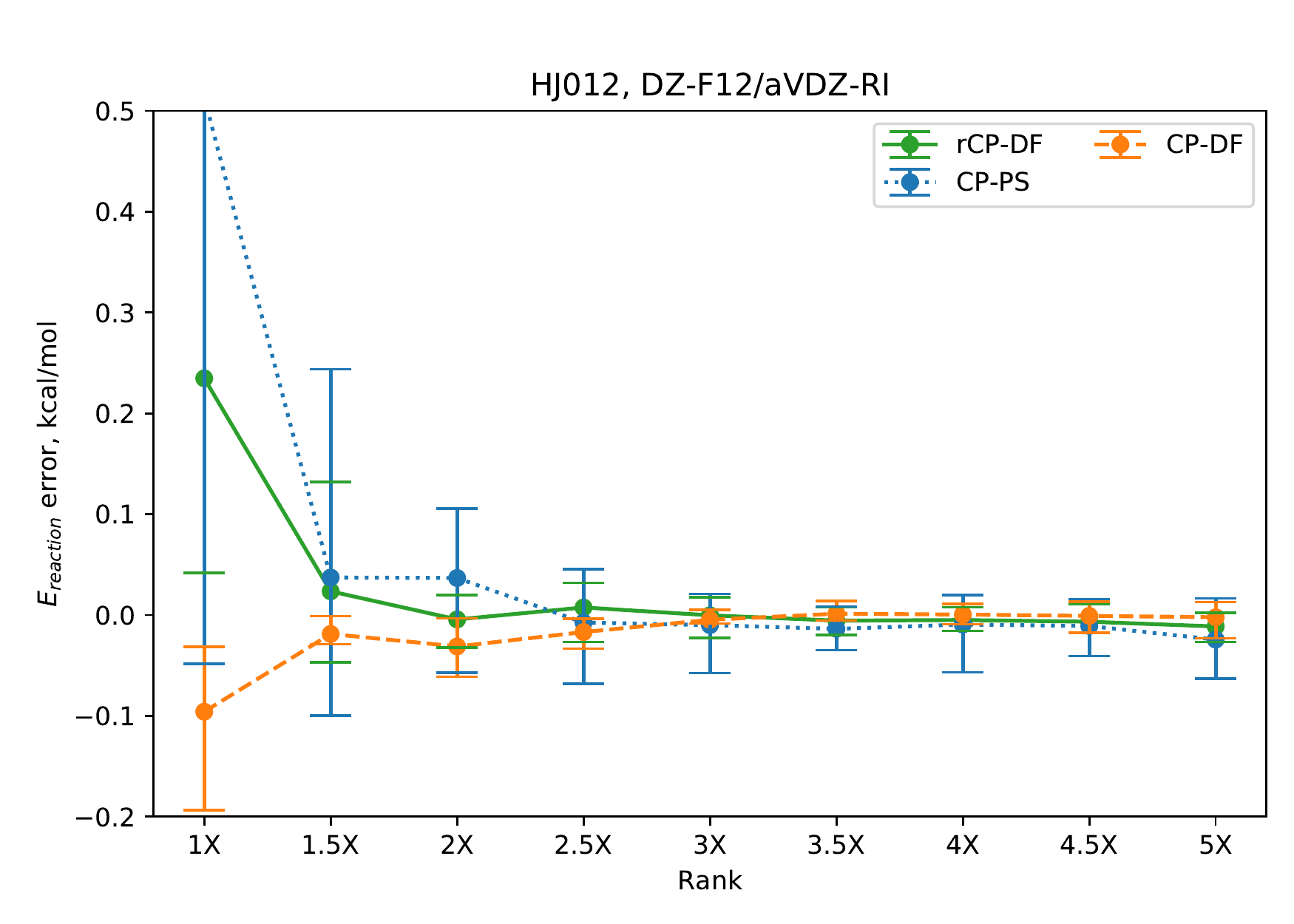}
      \caption{}
      \label{fig:RE-s-error-1e3}
    \end{subfigure}
    \caption{Mean unsigned (a) and signed (b) errors, respectively, in the CCSD reaction energies (kcal/mol) of the HJO12 dataset, relative to canonical CCSD, induced by the CP-DF, CP-PS or rCP-DF approximations to PPL vs CP rank $R$ (in units of the fitting basis, $X$). ALS precision fixed at $\epsilon = 10^{-3}$. The error bars denote the max/min errors.}
    \label{fig:RE}
  \end{figure}

        Next we examined the influence of the CP rank on the errors in chemical energy differences, rather than in absolute correlation energies. The unsigned and signed errors in the weak noncovalent binding energies of the S66/12 test set and in the HJO12 isogyric reaction energies are reported in \cref{fig:BE,fig:RE}, respectively. Because, compared to $R$, $\epsilon$ has a relatively small influence on $E_{\text{CCSD}}$, we have limited this assessment to using relatively loose ALS tolerances of $\epsilon = 10^{-3}$.\footnote{The corresponding results for a tighter ALS tolerance, $\epsilon = 10^{-4}$, are reported in the Supporting Information.} The target level of performance, defined here stringently as the maximum error of less than $0.1$ kcal/mol, is achieved with CP-DF and CP-PS when $R \geq 2X$. However, the use of rCP-DF allows us to attain the target accuracy with much smaller CP rank, $R \geq X$. For all relevant CP ranks, rCP-DF is at least an order of magnitude more accurate than CP-DF and CP-PS. As expected, the CP-PS errors are roughly a factor of 2 smaller than those due to CP-DF.

\begin{figure}
        \begin{subfigure}{0.45\textwidth}
            \includegraphics[width=0.9\linewidth]{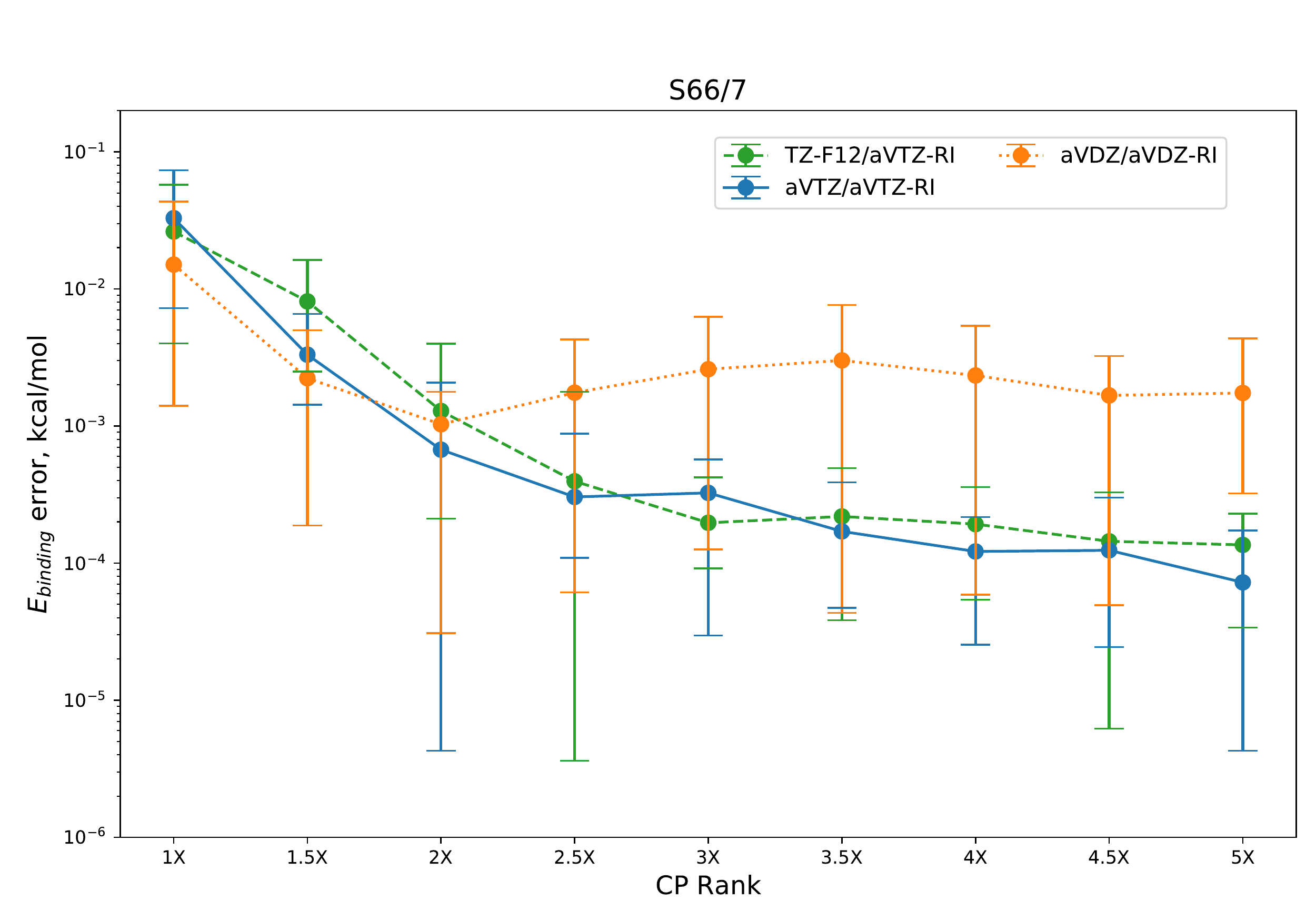}
            \caption{}
        \end{subfigure}
        \begin{subfigure}{0.45\textwidth}
            \includegraphics[width=0.9\linewidth]{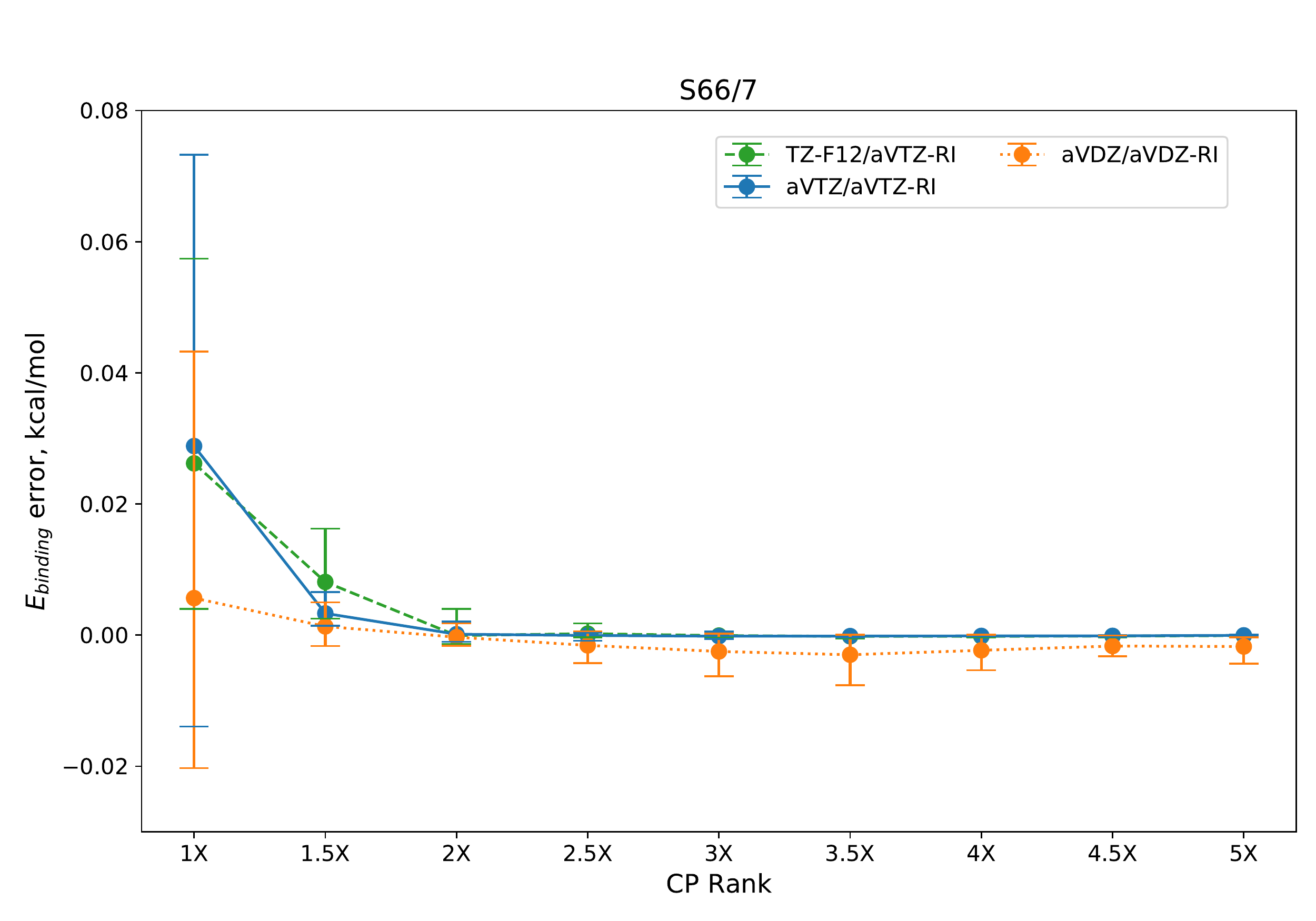}
            \caption{}
        \end{subfigure}
        \caption{Mean unsigned (a) and signed (b) errors, respectively, in the CCSD binding energies (kcal/mol) for the S66/7 dataset, relative to canonical CCSD, induced by the rCP-DF approximation to PPL vs CP rank $R$ (in units of the fitting basis, $X$) using 3 different basis sets, aVDZ/aVDZ-RI, aVTZ/aVTZ-RI and TZ-F12/aVTZ-RI. ALS precision fixed at $\epsilon = 10^{-3}$. The error bars denote the max/min errors. \label{fig:basis_sets}}
    \end{figure}

        The performance of the rCP-DF approximation to PPL is relatively insensitive to the basis set. Using the larger TZ-F12 OBS as well as the standard correlation-consistent aV{D,T}Z OBS does not appear to radically change the convergence trends, as illustrated in \cref{fig:basis_sets}.\footnote{A note of caution to the readers not familiar with the {D,T}Z-F12 basis sets: they are actually quite a bit larger than their conventional counterparts, and include even more diffuse Gaussians than the augmented correlation consistent basis sets} The errors in binding energies are small ($<0.1$ kcal/mol even with $R=X$) and rapidly decrease when $R$ is increased. The protracted convergence with the CP rank when using the aVDZ basis is somewhat puzzling, but is likely due to the need for tighter CP solver convergence for the smaller basis sets.
        
        It is instructive to compare the rCP-DF approximation for the PPL diagram with the best THC-based approach for the same, namely the least-squares THC(DF) method [LS-THC(DF)] and its {\em orbital-weighted} extension [W-LS-THC(DF)] developed by Parrish et al.\cite{Parrish2014}
        \cref{table:wat-hex} juxtaposes the maximum absolute and relative CCSD energy errors due to the rCP-DF and the THC PPL approximations for the 8 low-lying (\ce{H2O})$_{6}$ conformers. The same OBS/DFBS basis set pair, TZ/TZ-RI, was utilized for all computations. The rCP-DF approach used $R = 1.3 X$, whereas the corresponding LS-THC grid size corresponds to $R \approx 4 X$, i.e., roughly 3 times larger than used by our method. Although the absolute energies are most accurate with the W-LS-THC(DF) method of Parrish et al., the relative energies of the clusters are nearly as accurate with our method, despite its much smaller CP rank. Most importantly, the rCP-DF approach greatly outperforms its true THC counterpart, LS-THC(DF), again despite the much smaller CP rank. It is clear that the errors of the rCP-DF approach can be reduced further in the context of the CC methods by combining it with the orbital-weighting idea of Parrish et al.\cite{Parrish2014}
       
    \begin{table}[hb]
        \centering
        \begin{tabular}{cccccc}
        \hhline{======}
            & Maximum Absolute Error         &  Maximum Relative Error       \\ \hline
        rCP-DF  & 0.45                  & 0.036     \\
        LS-THC(DF)\cite{Parrish2014}  & 2.13                & 0.18      \\         W-LS-THC(DF)\cite{Parrish2014}  & 0.29                 & 0.03      \\ \hhline{======}
        \end{tabular}
        \caption{Maximum absolute and relative errors in valence TZ/TZ-RI DF-CCSD correlation energies (m$E_{\rm h}$) of 8 low-lying (\ce{H2O})${}_{6}$ clusters.\cite{Bates2011} For the rCP-DF approximation CP rank and ALS precision were fixed at $R = 1.3X$ and $\epsilon = 10^{-3}$, respectively. \label{table:wat-hex}}
    \end{table}
        
  \subsection{Cost Reduction vs DF-CCSD \label{sec:results_ccsd_cost}}
    Next we examined whether the stringent target errors in CCSD energies due to the rCP-DF PPL formulation can be attained along with demonstrated computational cost savings. 
    
    \begin{figure}[!t]
     \begin{subfigure}{0.47\textwidth}
      \includegraphics[width=0.9\linewidth]{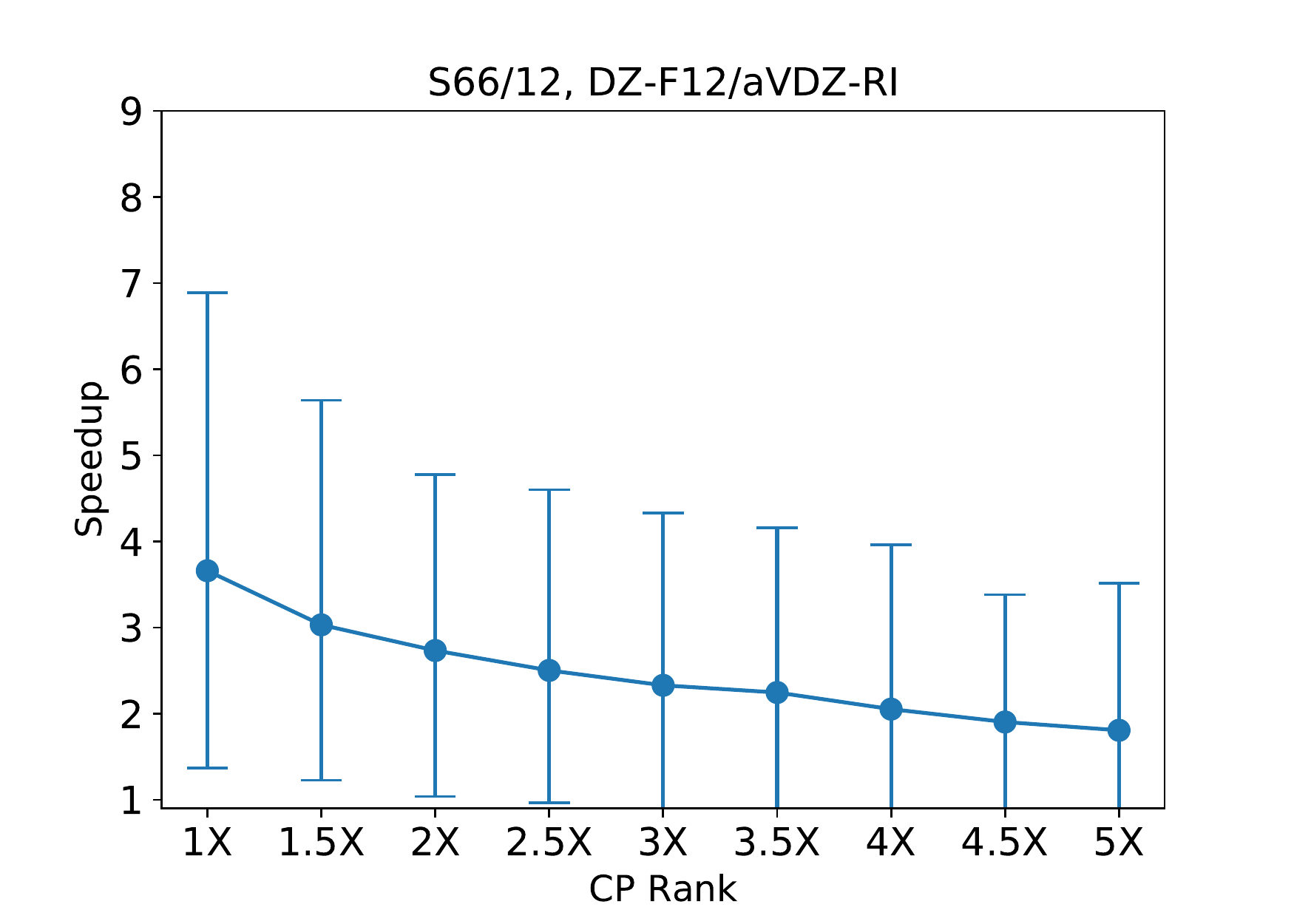}
      \caption{Average speedup (\cref{eqn:speedup}) of CCSD with rCP-DF-approximated PPL vs CP rank $R$ (in units of the fitting basis, $X$) for the S66/12 dataset. ALS precision fixed at $\epsilon = 10^{-3}$. The error bars denote the max/min speedup.}
      \label{fig:BEtime}
     \end{subfigure} \hfill
     \begin{subfigure}{0.47\textwidth}
      \includegraphics[width=0.9\linewidth]{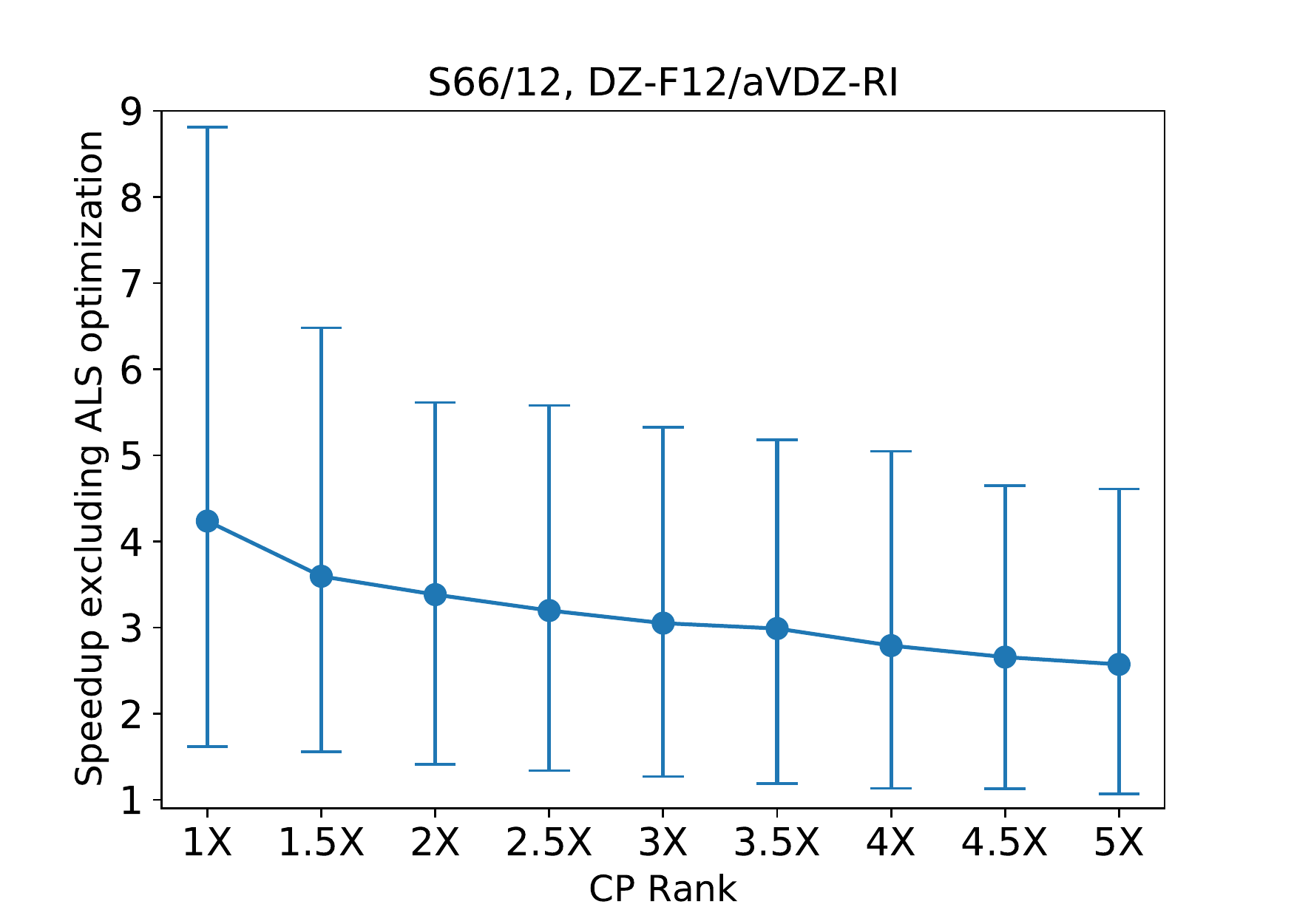}
      \caption{Average speedup (\cref{eqn:speedup}, \textbf{excluding} the cost of CP-ALS) of CCSD with rCP-DF-approximated PPL vs CP rank $R$ (in units of the fitting basis, $X$) for the S66/12 dataset. The error bars denote the max/min speedup.}
      \label{fig:Notime}
     \end{subfigure}
     \begin{subfigure}{0.47\textwidth}
       \includegraphics[width=0.9\linewidth]{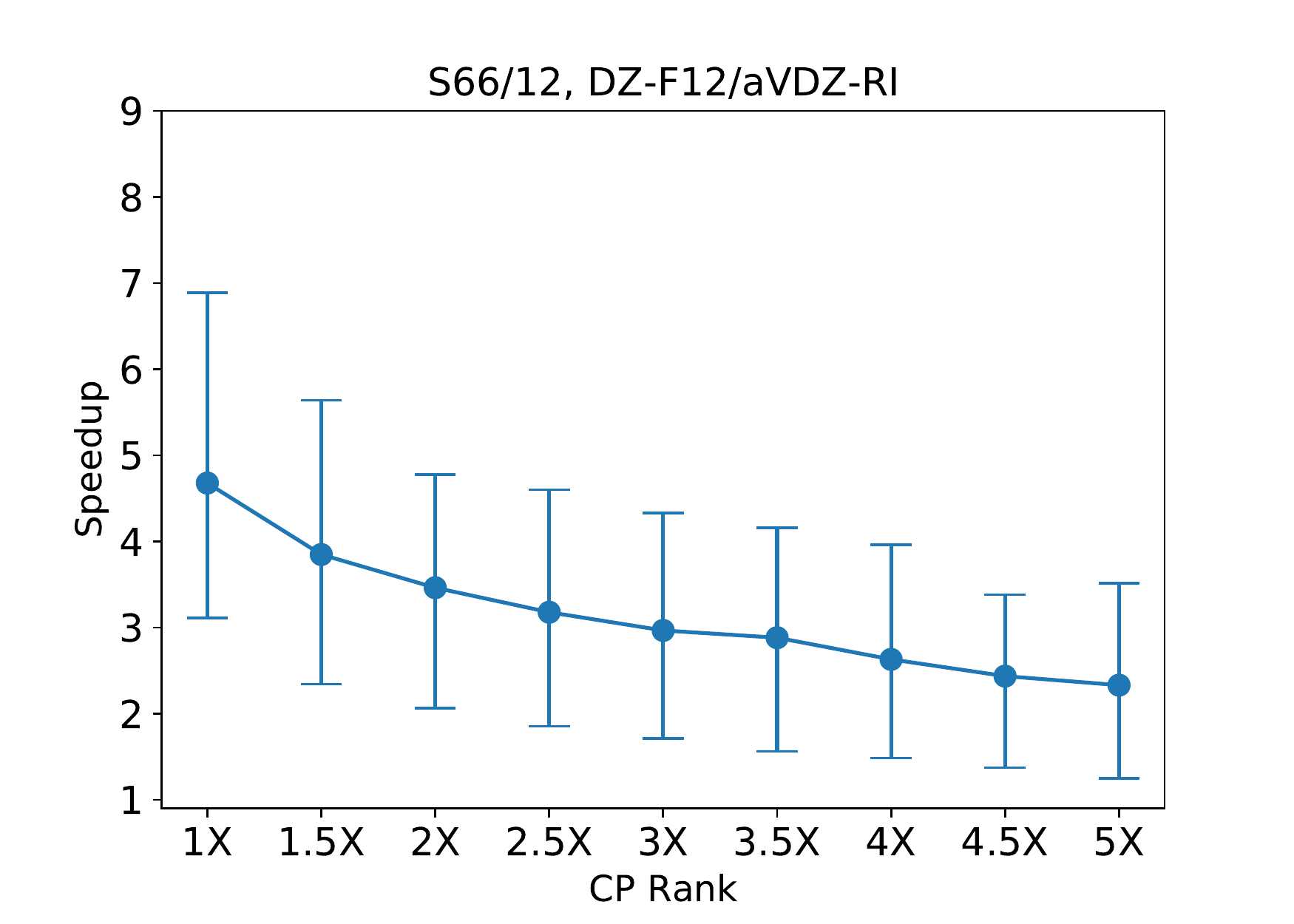}
      \caption{Average speedup (\cref{eqn:speedup}) of CCSD with rCP-DF-approximated PPL vs CP rank $R$ (in units of the fitting basis, $X$) for the 7 largest clusters in the S66/12 dataset. The error bars denote the max/min speedup.}
      \label{fig:Bigtime}
     \end{subfigure}
     \caption{}
    \end{figure}

    The observed speedups in the DF-CCSD computations due to the CP-based PPL reformulations are illustrated for the clusters in the S66/12 test set in \cref{fig:BEtime}. Just as in \Cref{sec:results_ccsd_errors_rank}, only $\epsilon = 10^{-3}$ are reported in the manuscript, with the $\epsilon = 10^{-4}$ results available in the Supporting Information. Significantly smaller average speedups were observed with $\epsilon = 10^{-4}$ compared to $\epsilon = 10^{-3}$, for the same CP rank. This suggests that the cost of ALS CP solver can increase dramatically with $\epsilon$, due to the increasing number of ALS iterations. To further illustrate this point, \cref{fig:Notime} demonstrates the speedups obtained by excluding the cost of ALS. We see that ALS has the most dramatic effect on cost when $\epsilon$ is tighter and $R$ is larger. 
    
    Unsurprisingly, ALS optimization had the greatest impact on the smallest molecules. \cref{fig:Bigtime} demonstrates that the speedup for the 7 largest clusters in the S66/12 set is significantly greater than the average speedup over the entire set and for all values of $R$. Since we found the energies relatively insensitive to the choice of $\epsilon$, we recommend the use of $\epsilon \approx 10^{-3}$ for all practical computations, unless extremely high target accuracy is sought.

  \begin{figure}
      \includegraphics[width=0.9\linewidth]{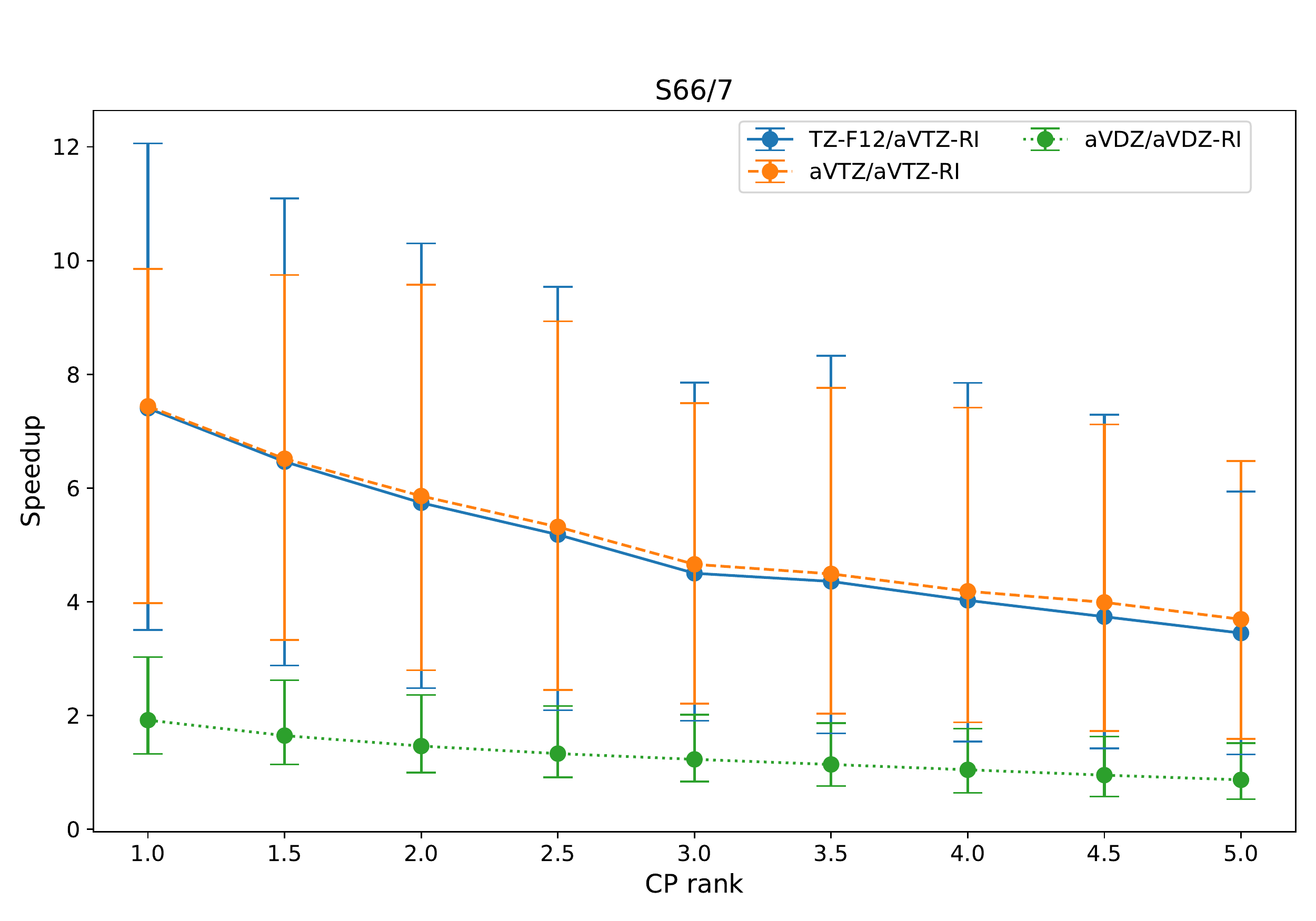}
      \caption{Average speedup (\cref{eqn:speedup}) of CCSD with rCP-DF-approximated PPL vs CP rank $R$ (in units of the fitting basis, $X$) for the S66/7 dataset. ALS precision fixed at $\epsilon = 10^{-3}$. The error bars denote the max/min speedup}
      \label{fig:time_other_basis}
  \end{figure}
    
    We further assessed the performance of the rCP-DF PPL approximation for the S66/7 dataset with 3 additional basis set pairs (\cref{fig:time_other_basis}). As one might expect, for larger basis sets, like TZ-F12 or aVTZ, the PPL diagram contributes significantly more to the cost of CCSD, hence even greater cost savings from rCP-DF are observed.
  \begin{figure}[!b]
      \includegraphics[width=0.9\linewidth]{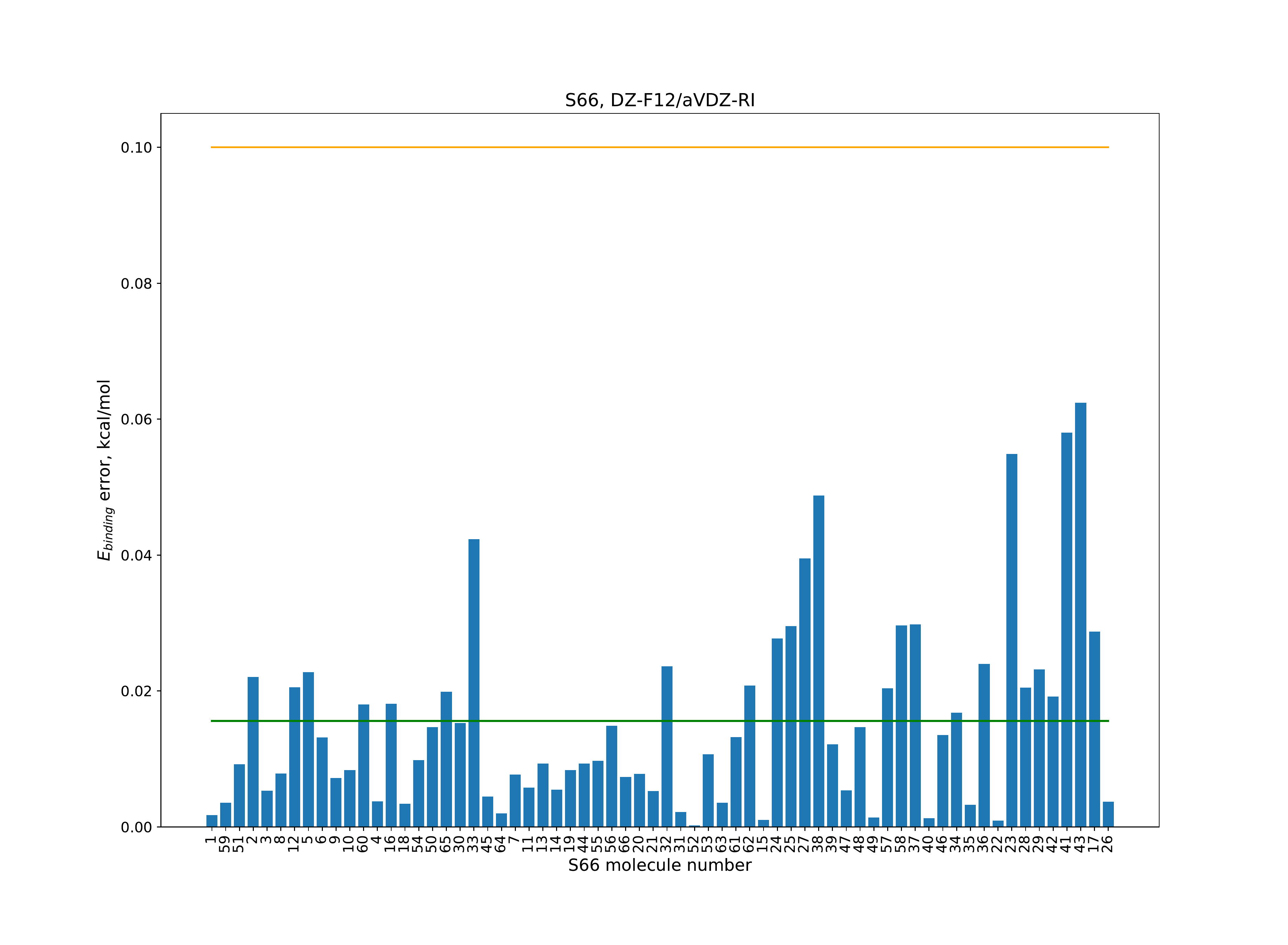}
      \caption{Unsigned errors in the S66 CCSD binding energies (kcal/mol), relative to canonical CCSD, induced by the rCP-DF approximation to PPL. CP rank and ALS precision are fixed at $R = 1.3X$ and $\epsilon = 10^{-3}$, respectively. Molecules ordered from smallest to largest number of occupied orbitals. The orange line is the target maximum error, $0.1$ kcal/mol, and the green line is the average error of the set.}
      \label{fig:s66error}
  \end{figure}
  \begin{figure}[!t]
      \includegraphics[width=0.9\linewidth]{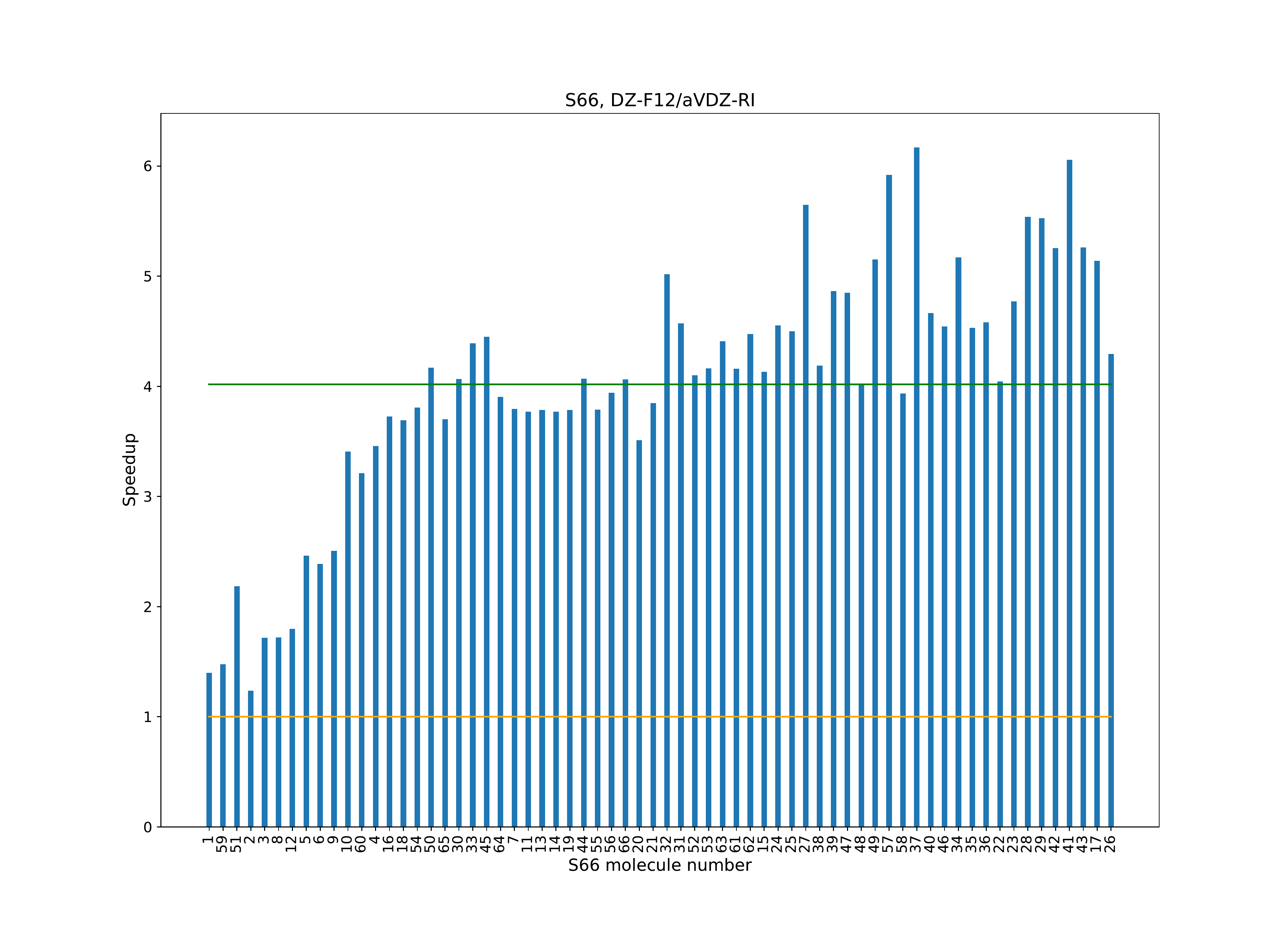}
      \caption{Speedup (\cref{eqn:speedup}) of CCSD with rCP-DF-approximated PPL for the entire S66 dataset. CP rank and ALS precision are fixed at $R = 1.3X$ and $\epsilon = 10^{-3}$, respectively. Molecules are ordered according to the number of occupied orbitals, from smallest to largest. The orange line represents no speedup over CCSD and the green line is average speedup of the set.}
      \label{fig:s66Time}
  \end{figure}
  
\begin{figure}[!t]
    \includegraphics[width=0.9\linewidth]{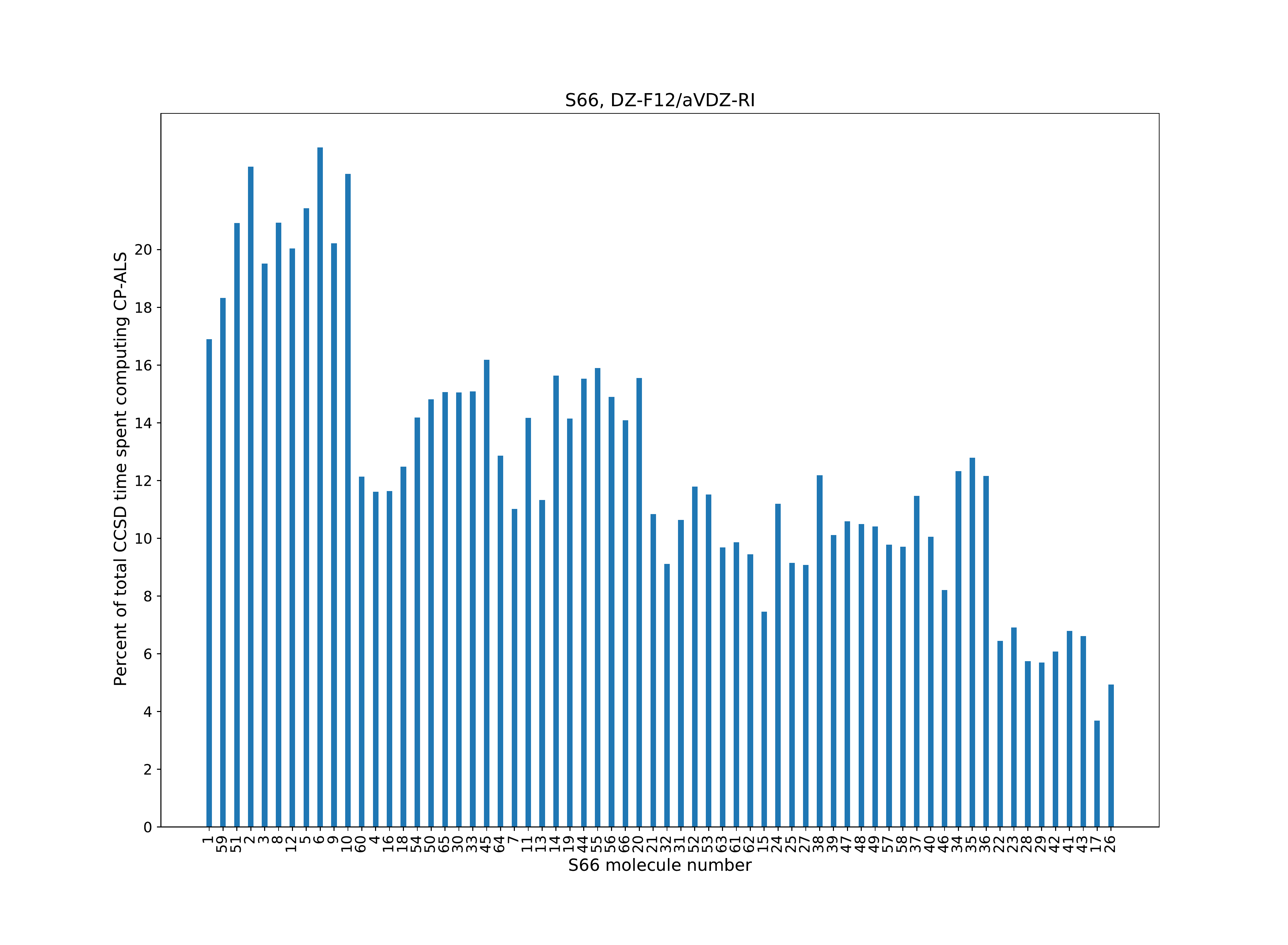}
    \caption{Percent of the total CCSD time spent in ALS for each cluster molecule in S66 dataset using rCP-DF with CP rank $R = 1.3X$ and ALS precision of $\epsilon = 10^{-3}$. Molecules are ordered according to the number of occupied orbitals, from smallest to largest.}
    \label{fig:percentS66}
  \end{figure}

    To further assess the performance of the rCP-DF PPL approximation, we computed the errors in CCSD binding energies for the entire S66 test set, using $R = 1.3X$ and $\epsilon = 10^{-3}$; the results are reported in \cref{fig:s66error,fig:s66Time}. For all systems, the errors introduced by rCP-DF are significantly less than $0.1$ kcal/mol, and the computational savings are realized for all systems, with the average speedups of $4$. This figure shows a clear trend: larger molecules benefit more from rCP-DF than smaller molecules. This trend is an artifact of the ALS optimization: as we increase the systems size, the cost of CCSD increases faster than the cost of the ALS and, thus, computing the ALS takes up a smaller percentage of the total CCSD time, as illustrated in \cref{fig:percentS66}. To note, although we only show speedup for the S66 cluster molecules, all of the dissociated cluster molecules also experienced a reduced cost over canonical DF-CCSD. The smallest dissociated molecule, a single water molecule, saw a cost reduction of a factor of 2.
    \begin{table}[hb]
        \label{table:h20}
        \centering
        \begin{tabular}{cccccc}
        \hhline{======}
            & $E_{\text{CCSD}}$         &  $D_e$        & $t_{\text{CCSD}}$     &  $t_{\text{PPL}}$     & $t_{\text{CP-ALS}}$ \\ \hline
        DF  & -5.02009                  & 182.47      & 1.36 $\times 10^{4}$  & 1.11 $\times 10^{4}$  & --- \\
        CP  & -5.02233                  & 182.44      & 3.47 $\times 10^{3}$  & 1.17 $\times 10^{3}$  & 2.32 $\times 10^{3}$ \\ \hline
            & \multicolumn{2}{c}{Error} & \multicolumn{2}{c}{Speedup}  \\ \hline
            &$\quad -1.41 \times 10^{-3}$  & $\quad2.81\times 10^{-2}$  & 3.92 & 9.46     \\ \hhline{======}
        \end{tabular}
        \caption{Valence CCSD correlation ($E_{\text{CCSD}}$, E$_{\text{h}}$) and dissociation energies ($D_e$, kcal/mol), the average per-iteration time spent in CCSD ($t_{\text{CCSD}}$, s) and its PPL contribution ($t_{\text{PPL}}$, s) for the (\ce{H2O})${}_{20}$ cluster. The total time of the CP ALS optimization is also reported ($t_{\text{CP-ALS}}$, s). CP rank and ALS precision are fixed at $R = 1.3X$ and $\epsilon = 10^{-3}$, respectively.}
    \end{table}
    To demonstrate the performance of the DF-CCSD method with the rCP-DF-approximated PPL term for a larger system, we used it to compute the binding energy of (\ce{H2O})${}_{20}$, with results reported in \cref{table:h20}. With the recommended values of $R$ and $\epsilon$, the cost of CCSD can be reduced by a factor of 3.8, with only a $\sim0.03$ kcal/mol impact on the binding energy.

\section{Summary and Perspective} \label{sec:discuss}
  In this work, we considered how {\em robust} (in the Dunlap sense\cite{Dunlap:2000iw}) approximation of tensor networks,
  in which the leading-order error due to the approximation of the network constituents is explicitly cancelled, can be used profitably
  to construct efficient factorizations of the 2-particle Coulomb interaction tensor. 
  We specifically considered tensor networks utilizing CP decomposition of order-3 tensors that arise from generalized square root factorizations of the Coulomb tensor, namely Cholesky and density fitting. Single use of the CP decomposition leads to a tensor network resembling the factorization in the well-known pseudospectral (PS) method, whereas double CP insertion leads to the tensor network topology of the tensor hypercontraction (THC) factorizations. Robust factorization combines these two base factorizations, resulting in a 1 to 2 order reduction of the error over either naive substitution scheme. Deeper analysis of the errors in the Coulomb interaction tensor revealed that the novel factorization, dubbed rCP-DF, corrects both errors resulting from the suboptimality of the CP factors as well as the errors due to deficient CP rank.
  
  As is also possible with the PS and THC factorizations, the rCP-DF factorization lowers the operation complexity of the cost-dominant PPL diagram in pair theories from \bigO{N^6} to \bigO{N^5}. Here we demonstrated in practice that the rCP-DF-approximated PPL can lower the practical cost of DF-CCSD even for systems with as few as 3 atoms. We make this claim because sufficiently small (on the thermal energy scale) errors can be achieved with a CP rank approximately equal to the rank of the density fitting basis itself; this hyperedge size requirement is substantially smaller than the requirements in previous PS and THC studies. For example, for the standard S66 and HJO12 benchmark sets of noncovalent interaction energetics and reaction energies, respectively, the use of such low CP rank induces {\em maximum} errors of only $\approx0.1$ kcal/mol. For the larger example of a 20-water cluster, the rCP-DF error in the dissociation energy was found to be only 0.03 kcal/mol.
  
  Although the complexity reduction due to the use of rCP-DF is very modest, the use of rCP-DF-PPL in the context of divide-and-conquer reduced-scaling CC approaches like FMO,\cite{Kitaura1999} CIM,\cite{Li:2009CIM} DEC,\cite{Kristensen2011,Kjaergaard:2017gb} and others\cite{Friedrich2007}, might be beneficial to reduce the cost of the fragment computation.
  
  The proposed robust tensor factorization of the Coulomb interaction, clearly, can be improved further, as well as applied in other contexts. Some of the promising ideas are listed here:
  \begin{itemize}
  \item This particular robust CP-based factorization, which we consider here, utilized the density-fitting-based generalized square root factorization of the Coulomb tensor. Though, it should be trivial to apply the factorization to other square-root factorizations, such as the (pivoted) Cholesky.
  \item Although we only considered algebraic CP decomposition of the square root factor, it should be possible to use the idea in the context of quadrature-based factorization, such as PS, COSX, and least-squares THC. For example, robust LS-THC should allow for the use of smaller grids than currently possible (the juxtaposition of the rCP-DF and LS-THC(DF) performance in \cref{sec:results_ccsd_errors_rank}, albeit limited, suggests that grid size reductions of a factor of 3 or more are realistic). Robust factorization should also simplify formulation of analytic gradients.
  \item A combination with other ideas such as the use of orbital-biasing explored in LS-THC-based coupled-cluster\cite{Parrish2014} and the use of frozen natural orbitals should be beneficial.
  \item The efficiency of the CP solver can be greatly improved via the use of gradient-based techniques.
  \end{itemize}
  Work along some of these directions is underway.
  
\begin{acknowledgement}
This work was supported by the U.S. National Science Foundation (awards 1550456 and 1800348). We also acknowledge Advanced Research Computing at Virginia Tech (www.arc.vt.edu) for providing computational resources and technical support that have contributed to the results reported within this paper.
\end{acknowledgement}

~\\
{\bf {\Large Supporting Information}}\\
Results with $\epsilon=10^{-4}$ and the ALS algorithm for computing the optimal (for fixed rank) CP-DF approximation of the Coulomb tensor.

\bibliography{cp3-ccsd-refs.bib}

\end{document}